\documentclass{aa}   
\usepackage[utf8]{inputenc}
\usepackage[T1]{fontenc}
\usepackage{graphicx}
\usepackage{fancyhdr}
\usepackage{nomencl}
\usepackage{setspace}
\usepackage{enumitem}
\usepackage{etoolbox}
\usepackage{amsmath}
\usepackage{amssymb}
\usepackage{physics}
\usepackage{mathrsfs}
\usepackage{natbib}
\usepackage[colorlinks]{hyperref}
\usepackage{algpseudocode}
\usepackage{algorithm}
\usepackage{booktabs}
\usepackage{subfigure}
\usepackage{dsfont}
\usepackage{xcolor}
\usepackage{stmaryrd}
\usepackage{diagbox}
\usepackage{mathtools}
\usepackage{mdframed}
\usepackage{bm}
\usepackage{blindtext}
\usepackage[export]{adjustbox} 
\usepackage{multirow}
\usepackage{makecell}

\graphicspath{{./Images/}} 

\newcommand*{\tran}{^{\mkern-1.5mu\mathsf{T}}}

\definecolor{myblue}{RGB}{12, 12, 158}
\definecolor{myred}{RGB}{158, 19, 22}
\definecolor{myorange}{RGB}{245, 150, 12}
\definecolor{mygreen}{RGB}{26, 148, 49}
\definecolor{Prune}{RGB}{99,0,60}
\definecolor{Purple}{RGB}{75, 0, 130}
\definecolor{Pink}{RGB}{255, 105, 180}

\definecolor{deepskyblue}{RGB}{0, 191,255}
\definecolor{limegreen}{RGB}{50, 205, 50}

\hypersetup{
    citecolor=[rgb]{0.04, 0.20, 0.58},  
    linkcolor=[rgb]{0.86, 0.078, 0.235},
    urlcolor=[rgb]{0,0,1}
}

\definecolor{void}{RGB}{63,96,174}
\definecolor{wall}{RGB}{109,179,136}
\definecolor{filament}{RGB}{231,133,50}
\definecolor{node}{RGB}{217,33,32}
\definecolor{combination}{RGB}{83,158,182}

\bibpunct{(}{)}{;}{a}{}{,} 

\makeatletter
\renewcommand*\aa@pageof{, page \thepage{} of \pageref*{LastPage}}

\begin{document}

\title{Cosmology with cosmic web environments}
\subtitle{II. Redshift-space auto and cross power spectra}

\author{Tony Bonnaire\inst{1}, Joseph Kuruvilla\inst{2}, Nabila Aghanim\inst{2}, Aurélien Decelle\inst{3,4}}
\authorrunning{T. Bonnaire et al.}
\titlerunning{Cosmology with cosmic web environments II.}
\institute{
            Laboratoire de Physique de l’École normale supérieure, ENS, Université PSL, CNRS, Sorbonne Université, Université Paris Cité, F-75005 Paris, France \\
            E-mail: \href{mailto:tony.bonnaire@ens.fr}{tony.bonnaire@ens.fr}
            \and
            Université Paris-Saclay, CNRS, Institut d'Astrophysique Spatiale, 91405, Orsay, France
            \and
            Université Paris-Saclay, TAU team INRIA Saclay, CNRS, Laboratoire Interdisciplinaire des Sciences du Numérique, 91190, Gif-sur-Yvette, France
            \and
            Departamento de Física Teórica I, Universidad Complutense, 28040 Madrid, Spain
        }
\abstract{
Degeneracies among parameters of the cosmological model are known to drastically limit the information contained in the matter distribution. In the first paper of this series, we shown that the cosmic web environments; namely the voids, walls, filaments and nodes; can be used as a leverage to improve the real-space constraints on a set of six cosmological parameters, including the summed neutrino mass.

Following-upon these results, we propose to study the achievable constraints of environment-dependent power spectra in redshift space where the velocities add up information to the standard two-point statistics by breaking the isotropy of the matter density field. A Fisher analysis based on a set of thousands of Quijote simulations allows us to conclude that the combination of power spectra computed in the several cosmic web environments is able to break some degeneracies. Compared to the matter monopole and quadrupole information alone, the combination of environment-dependent spectra tightens down the constraints on key parameters like the matter density or the summed neutrino mass by up to a factor of $5.5$. Additionally, while the information contained in the matter statistic quickly saturates at mildly non-linear scales in redshift space, the combination of power spectra in the environments appears as a goldmine of information able to improve the constraints at all the studied scales from $0.1$ to $0.5$ $h$/Mpc and suggests that further improvements are reachable at even finer scales.
}


\defcitealias{DESI16}{DESI~Collaboration~2016}

\maketitle


\section{Introduction} \label{sect:introduction}

The spatial distribution of matter in the universe is a powerful probe to constrain cosmological parameters \citep[see e.g.][]{Totsuji1969, Peacock2001, Eisenstein2005, Mandelbaum2013, Alam2016, Hildebrandt2017, Ivanov2020}, test theory of gravity \citep{Alam2017, Jullo2019, Blake2020, Alam2020} or improve our understanding of dark matter and dark energy \citep{Abbott2018, Drlica-Wagner2019}. All these reasons have led numerous collaborations over the past decades to initiate a race to map the largest number of galaxies in the sky with the highest possible accuracy. The stage IV of these galaxy redshift surveys, with notably the Dark Energy Spectroscopic Instrument\footnote{\url{https://www.desi.lbl.gov/}} \citep{DESICollaboration2016}, Euclid\footnote{\url{https://www.euclid-ec.org/}} \citep{Leureijs2011} and the Nancy Grace Roman space telescope\footnote{\url{https://roman.gsfc.nasa.gov/}} \citep{WFIRST15}, promises an unprecedented amount of data enabling, among other things, the most accurate estimate of cosmological parameters.
These expected observational achievements must however be accompanied with the appropriate theoretical and numerical advances. Even though largely used in most analyses because of its simplicity, it is for instance well-known that the power spectrum (or equivalently the two-point correlation function) is not a sufficient summary statistic when it comes to non-Gaussian fields like the distribution of matter in the Universe at late time. Another caveats of using the matter power spectrum for cosmological analyses is the degeneracies among parameters of the $\Lambda$ cold dark matter ($\Lambda$CDM) model that have a similar impact over a wide range of scales, which limits the information one can extract from it.

Over the past years, many works investigated several directions of research to fight the degeneracies and go beyond simple power spectrum analyses in order to optimally exploit the matter distribution. These include information coming from higher-order statistics \citep{Hahn2020, Hahn2021, Gualdi2021, Philcox2022}, velocities \citep{Mueller2015, Kuruvilla2021, Kuruvilla2021a}, marked power spectrum \citep{Massara2021, Massara2022}, wavelet scattering transform \citep{Allys2020, Cheng2020, Cheng2021, Valogiannis2021, Valogiannis2022, Eickenberg2022, Wang2022, Wang2022a}, split densities \citep{Uhlemann2020, Paillas2021, Paillas2022}, partial or total cosmological environments \citep{Kreisch2021, Bayer2021, Bonnaire2021c, Woodfinden2022}, or from the minimum spanning tree \citep{Naidoo2019, Naidoo2021}. While the recent literature abounds of analyses and forecasts aiming at identifying the optimal statistic in real space, only a few address the problem in the space the data are actually collected, i.e. the redshift space. It is nonetheless well-known that the additional information carried by velocities coming into play in this space is non-negligible and leads to considerable improvement of the cosmological constraints derived from the standard matter statistics.

The real-space results presented in \cite{Bonnaire2021c} demonstrated the ability of simple two-point statistics derived in the separate cosmic web environments (voids, walls, filaments, and nodes) to break key degeneracies in the cosmological model, consequently enabling the improvement of the constraints for all the parameters over those derived from the matter power spectrum.
In this work, we extend our analysis of the constraints drawn by the combination of power spectra computed in the several environments when considering redshift-space distortions (RSD). In particular, we investigate how the combination of environment-dependant spectra in this space can improve the cosmological constraints and how they compare to the real-space case without correcting for the distortions by any means. To this end, we perform a Fisher forecast of the constraints using numerous simulations with varying cosmological parameters to assess numerically the information contained in our statistics on the cosmological parameters.
The paper is organised such that Sect. \ref{sect:data} presents the Quijote suite of simulations and the methodology used to extract the cosmic-web environments while Sect. \ref{sect:redshift_space} introduces the power spectra estimation in redshift space and in the cosmic web environments. Section \ref{sect:constraints} then presents the forecast constraints in redshift-space but also compares with the real space findings reported in \cite{Bonnaire2021c}. Finally, Sect. \ref{sect:disc_ccl}, in addition to wrap up and conclude, also discusses the obtained results in view of the recent literature and gives perspectives for future steps towards use of such analyses in observational setups.


\section{Data \& methodology} \label{sect:data}

\subsection{The Quijote simulations}

Our analysis relies on the Quijote\footnote{\url{https://quijote-simulations.readthedocs.io/en/latest/}} \citep{Villaescusa-Navarro2019} suite providing a set of $44\,100$ $N$-body simulations. With realisations of more than $7\,000$ cosmological models, this large suite of simulations was precisely designed to perform statistical analyses and train machine learning algorithms. Initialised with the second-order Lagrangian perturbation theory (or Zel'dovich approximation in case of massive neutrino simulations), $N=512^3$ dark matter particles (and $512^3$ neutrinos in case there are) are evolved forward in time in an $L=1$ Gpc/$h$ size box from redshift $z=127$ to $z=0$. There are $15\,000$ simulations available at the fiducial cosmology, assumed to be a flat $\Lambda$CDM cosmology with parameters $\Omega_\mathrm{m} = 0.3175$, $\Omega_\mathrm{b} = 0.049$, $h = 0.6711$, $n_\mathrm{s} = 0.9624$, $\sigma_\mathrm{8} = 0.834$ and $M_\nu = 0$, with $M_\nu$ the summed neutrino mass. For each cosmological parameter is then individually computed $500$ simulations with a fixed increase and decrease being $\dd \Omega_\mathrm{m} = \pm0.010$, $\dd \Omega_\mathrm{b} = \pm0.002$, $\dd h = \pm0.020$, $\dd n_\mathrm{s} = \pm0.020$, and $\dd \sigma_8 = \pm0.015$. Because $M_\nu$ is a positive quantity, the Quijote suite provides four positive variations of this parameter with $M_\nu = \sum m_\nu = \{0.1, 0.2, 0.4\}$ eV.

\subsection{Cosmic web classification}  \label{subsect:method_tweb}

Among the many possible definitions of the cosmic web environments proposed in the literature \citep[see][to name only a few]{Stoica2007, AragonCalvo2010, Nexus, DisperseTheory, Tempel2014, Bonnaire2020, Bonnaire2021b}, we resort to an implementation of the T-Web algorithm \citep{Hahn2007, Forero-Romero2009}. We summarise the general steps of the classification hereafter, but refer to \cite{Bonnaire2021c} for more details about the procedure and discussions about the parameters and their effects.
We first start by computing a smooth density field $\rho(\bm{x})$ based on the discrete set of particles by means of a $B$-spline interpolation of order four over an $N_\mathrm{g}^3 = 360^3$ three-dimensional grid. The T-Web formulation relies on the tidal tensor $\bm{T}(\bm{x})$ that is the second derivative of the gravitational potential, itself computed from $\rho(\bm{x})$ and the Poisson equation. Each of the grid cell $\bm{x}$ is classified in either void, wall, filament or node depending on the eigenvalues of the tidal tensor. More precisely, a cell $\bm{x}$ is in a void if the three eigenvalues are below $\lambda_\mathrm{th}$, in a wall if only two are, in a filament if only one is, and in a node if none is below the threshold.
From the classification of the $N_\mathrm{g}^3$ cells in the gridded density field, we propagate the environments at the particle level, enabling the computation of four individual overdensity fields $\delta_\mathrm{v}$, $\delta_\mathrm{w}$, $\delta_\mathrm{f}$ and $\delta_\mathrm{n}$ where the subscript respectively refer to void, wall, filament and node. They are all linked to the full matter overdensity field $\delta_\mathrm{m}$ by the linear combination $\delta_\mathrm{m} = \sum_{\alpha \in \{\mathrm{v}, \mathrm{w}, \mathrm{f}, \mathrm{n}\}} f_\alpha \delta_\alpha$, where $f_\alpha = N_\alpha / N$ denotes the mass fraction\footnote{Since all the particles have the same mass in the simulation, the ratio between $N_\alpha$, the number of particles in the environment $\alpha$ and the total number of particles $N$ gives the mass fraction.} of the environment $\alpha$.

The classification scheme is relying on three parameters. The first two are related to the coarseness of the smooth gravitational potential estimation: $N_\mathrm{g}^3$, the number of grid cells and $\sigma_\mathcal{N}$, the scale of the Gaussian with which the potential is being smoothed before performing the classification. The last parameter, $\lambda_\mathrm{th}$, corresponds to the threshold of the classification rules used for the amplitudes of eigenvalues of the tidal tensor.
In \cite{Bonnaire2021c}, we explored the impact of the parameters and ultimately set $N_\mathrm{g}^3 = 360^3$ cells and a value of $\sigma_\mathcal{N} = 2$ Mpc/$h$ for the smoothing scale which yields an effective smoothing of $3.4$ Mpc/$h$. The threshold eigenvalue is set to $\lambda_\mathrm{th} = 0.3$ to obtain a cosmic web in which the voids starts to percolate and is consistent with the use of such a methodology in the literature \citep[see e.g.,][]{Hahn2007, Martizzi2018, Libeskind2017, Bonnaire2021c}.

\section{Redshift-space} \label{sect:redshift_space}

\subsection{Redshift-space distortions}

When dealing with observational data, the redshift is used as a measure of the distance. However, this quantity is the combination of the proper motion of the source due to its peculiar velocity, $\bm{v}$, and the expanding universe.
Considering $\bm{x}_\mathrm{r}$ the position of a source in the comoving space, $\bm{x}_s$ its position in the redshift space, and $\hat{\bf{n}}$ a unit vector in the line of sight (LoS) direction, it holds the mapping relation
\begin{equation} \label{eq:RSD}
    \bm{x}_\mathrm{s} = \bm{x}_\mathrm{r} + \bigg( \frac{\bm{v} \cdot \hat{\bf{n}}}{aH} \bigg) \hat{\bf{n}}\,,
\end{equation}
with $a$ the scale factor and $H$ the Hubble constant. This shift thus distorts the spatial distribution of the tracers, and is known as ``redshift-space distortion'' (RSD), which leads to two dominant effects. First, the Finger-of-God (FoG) effect \citep{Jackson1972} makes overdense clustered regions appear elongated in the LoS direction due to the high velocity of sources. Second, on larger scales, a squashing effect of dense regions along the LoS is observed and known as the ``Kaiser effect'' \citep{Kaiser1987}.

While \cite{Bonnaire2021c} focused on the information content of power spectra in cosmic web environments computed in real space, the current work proposes to investigate the information in redshift-space, hence being closer to observations. To do so, we mimic the effect of RSDs in each individual simulation by displacing all particles (dark matter particles and neutrinos if any) according to Eq.~\eqref{eq:RSD} along one Cartesian axis of the box hence assuming the plane-parallel approximation.

\subsection{Power spectra estimation in redshift-space} \label{subsect:method_power_spectra}

In a general manner, the cross-power spectrum is defined as the covariance of Fourier transformed overdensity fields $\delta_\alpha$ and $\delta_\beta$ and is given by
\begin{equation}  \label{eq:PS_redshift}
    P_{\mathrm{s}, \alpha\beta}^\ell(k) = \frac{2\ell + 1}{2} \int_{-1}^1 P_{\alpha\beta}(k, \mu) \mathcal{L}_\ell(\mu) \dd \mu,
\end{equation}
with $\mu = \bm{k} \cdot \hat{\bf{n}} / k$, the angle with the line of sight, $P_{\alpha\beta}(k,\mu)$ the 2D power spectrum obtained by binning both in $k$ and $\mu$, and $\mathcal{L}_\ell$ the Legendre polynomials. In real space, the isotropy of the density field implies all the $\ell > 0$ terms to cancel out leaving the monopole alone carry all the information. In redshift space, the peculiar velocities of particles induce a dependence of the power spectrum with the line-of-sight hence leading to the breaking of the density field isotropy and consequently spreads the power over multiple poles $\ell$.
Throughout the paper, we employ the notation $P_{\alpha\beta}(k)$ to denote cross- or auto-spectra monopole in real space and keep the subscript $\mathrm{s}$ to denote statistics computed in redshift-space, together with the superscript $\ell$ referring to the pole.
To characterise spectra in redshift-space, we rely on the three first non-zero multipoles, namely $P^{\ell=0}_{\mathrm{s}, \alpha\beta}(k), P^{\ell=2}_{\mathrm{s}, \alpha\beta}(k), P^{\ell=4}_{\mathrm{s}, \alpha\beta}(k)$, respectively called monopole, quadrupole and hexadecapole. These $\ell \leq 4$ orders are the only non-vanishing moments in the linear approximation of the distortions and encode the full 2D information at linear scales \citep{Kaiser1987}. The corresponding Legendre polynomials are
\begin{equation}
    \mathcal{L}_\ell(\mu) =
    \begin{cases}
        1 & \text{if} \; \ell=0, \\
        \left(3 \mu^2 - 1\right)/2 & \text{if} \; \ell=2, \\
        \left(35 \mu^4 - 30 \mu^2 + 3\right)/8 & \text{if} \; \ell=4.
    \end{cases}
\end{equation}
Similarly to \cite{Bonnaire2021c}, we deconvolve the fields $\delta_\alpha$ before estimating the power spectra to remove biasing effect in the estimation introduced by the smoothing of the fourth-order $B$-spline interpolation to obtain the density fields. To additionally be robust to any bias provoked by aliasing effects, the maximum Fourier bin, that we set at $k_\mathrm{max} = 0.5$ $h$/Mpc by default, is below half of the Nyquist frequency defined as $k_\mathrm{Nyq} = \pi N_\mathrm{g} / L= 0.57$ $h$/Mpc.

\begin{figure}
    \centering
    \includegraphics[width=1\linewidth]{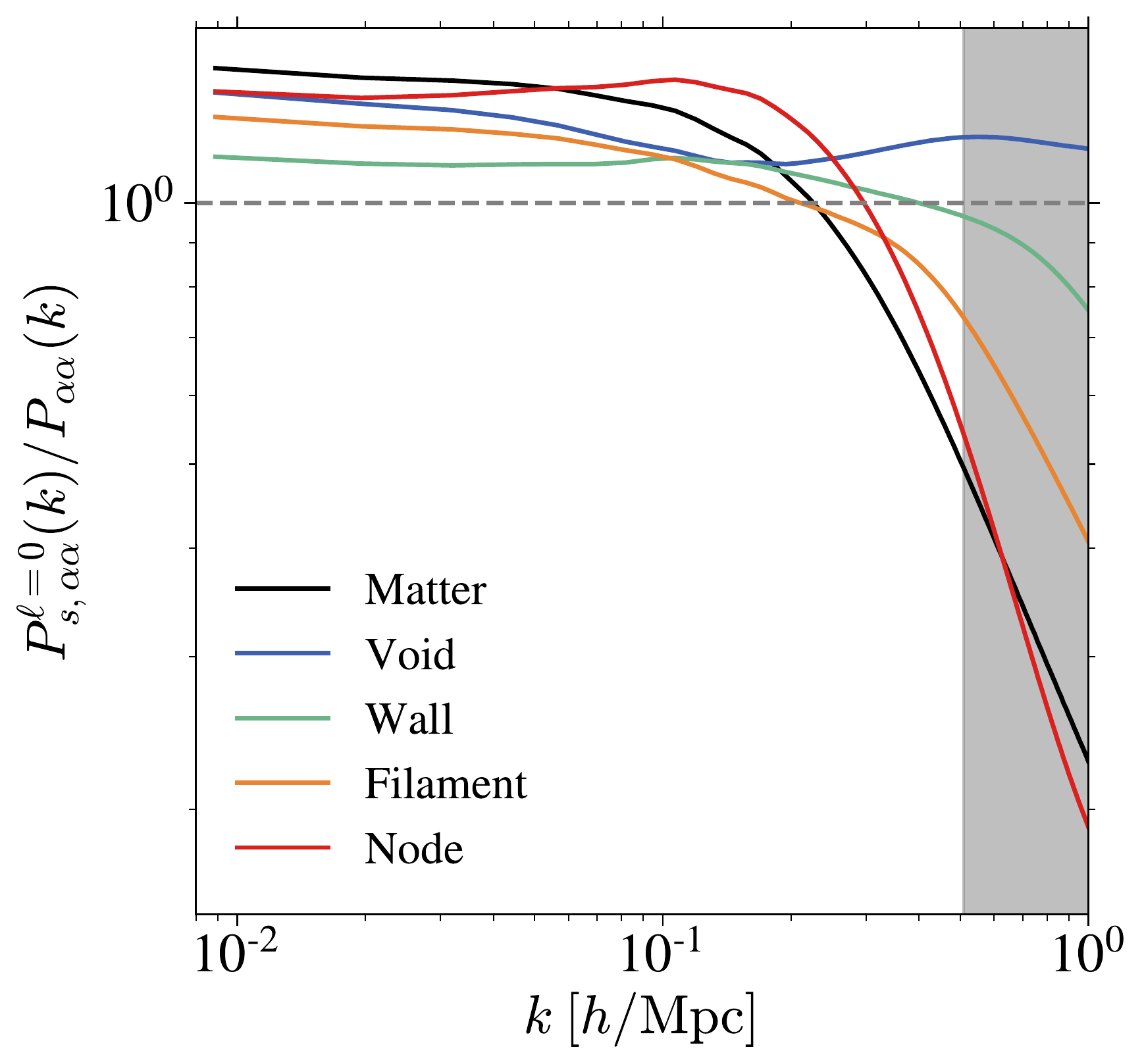}
    \caption{Ratio between the redshift-space and real space monopoles computed in the cosmic web environments. The grey shaded region shows the scales $k>0.5$ $h$/Mpc unused in this analysis.}
    \label{fig:monopoles_real_rsd}
\end{figure}

\subsection{Cosmic web environments and redshift-space} \label{subsect:cw_redshift}

\begin{figure*}
    \centering
    \includegraphics[width=.49\linewidth]{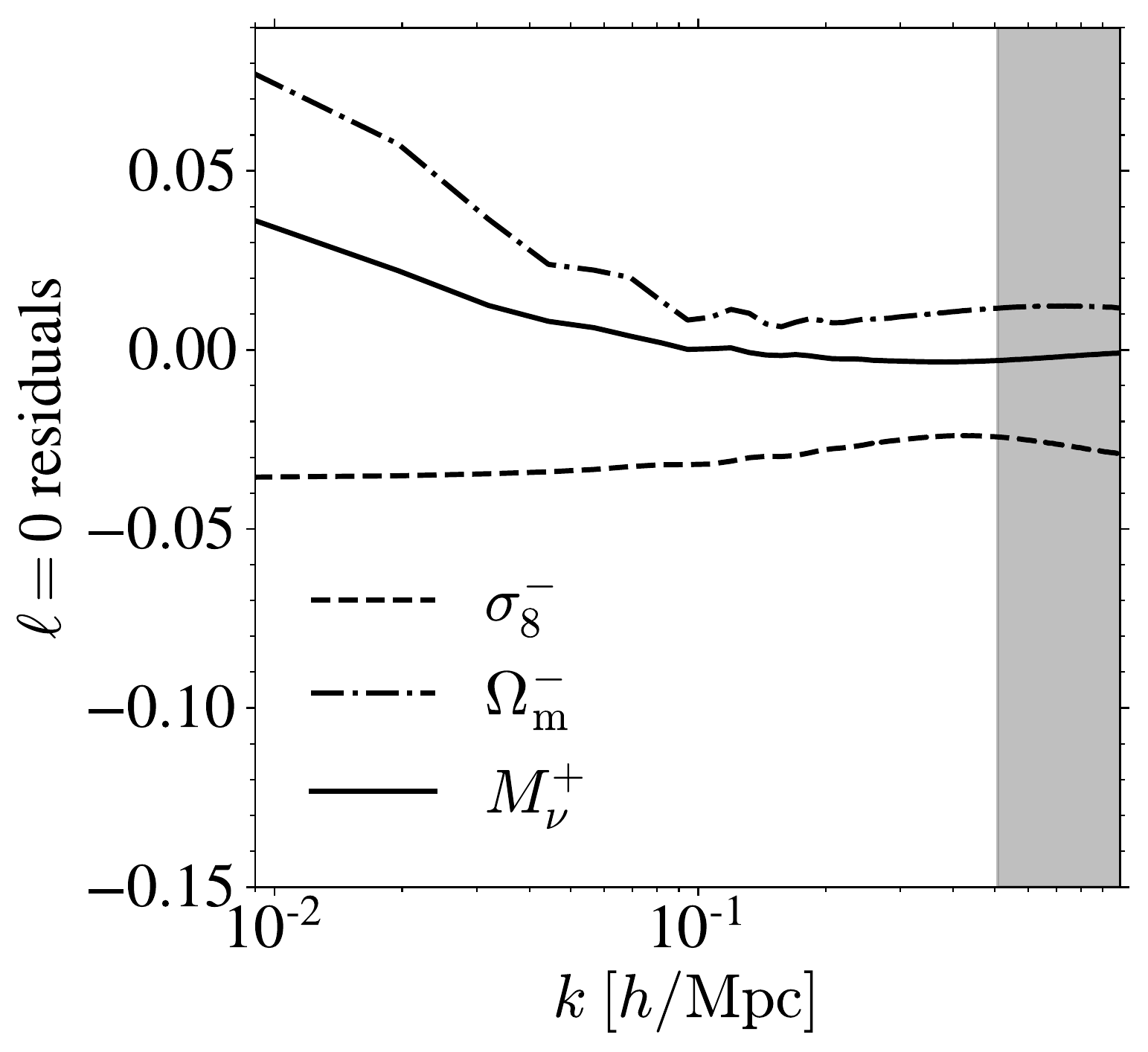}
    \includegraphics[width=.49\linewidth]{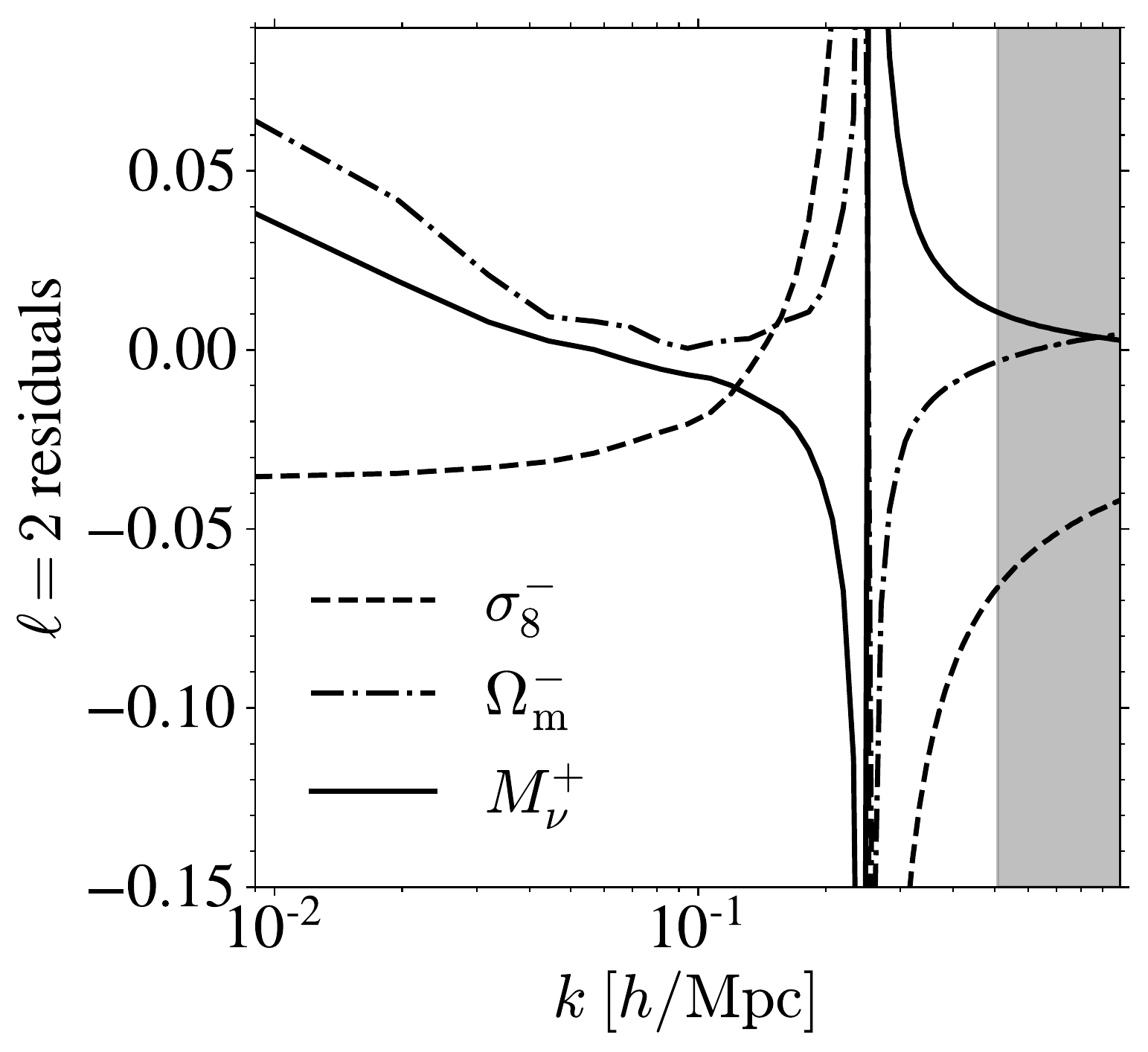}
    \\
    \includegraphics[width=.49\linewidth]{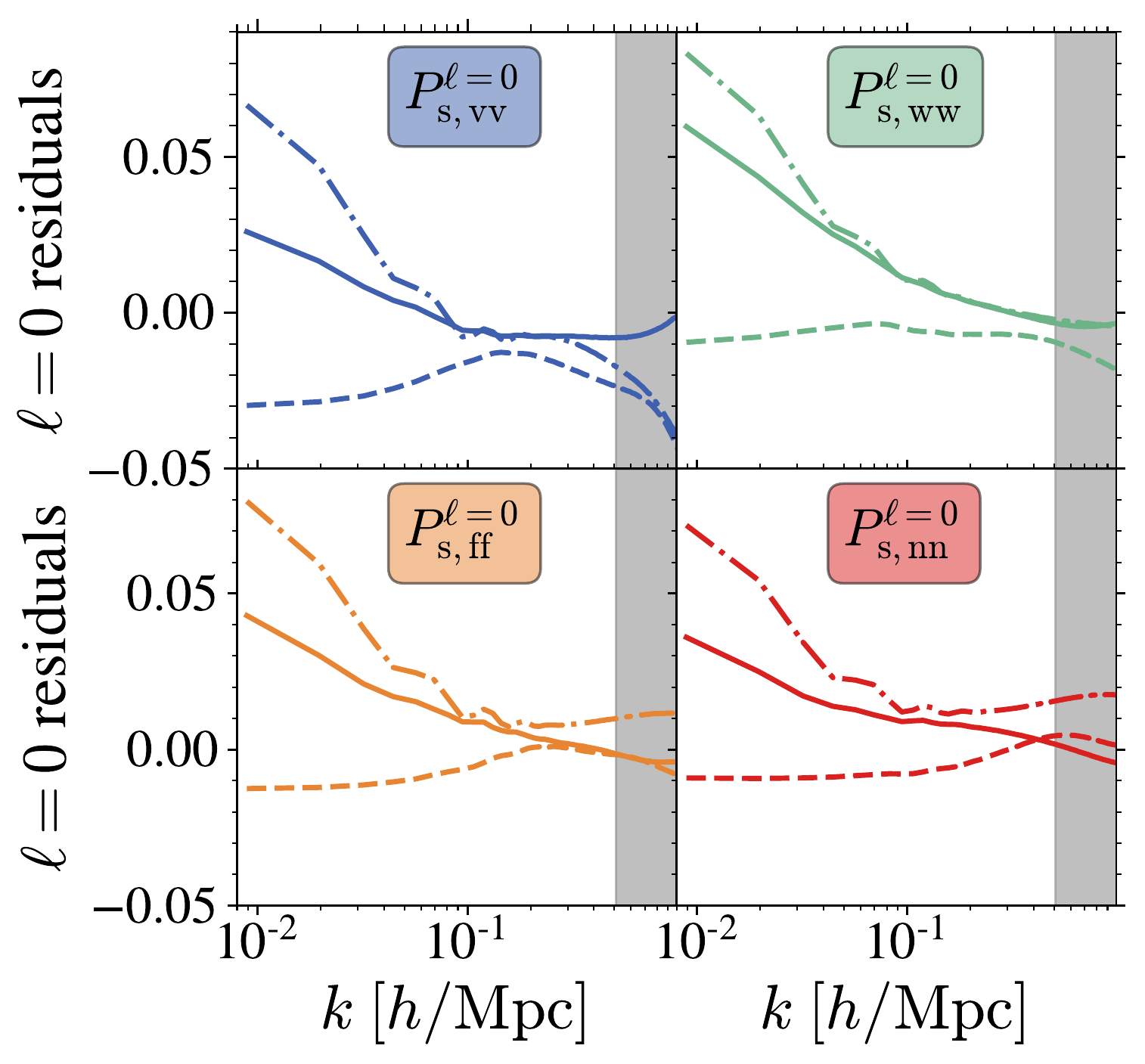}
    \includegraphics[width=.49\linewidth]{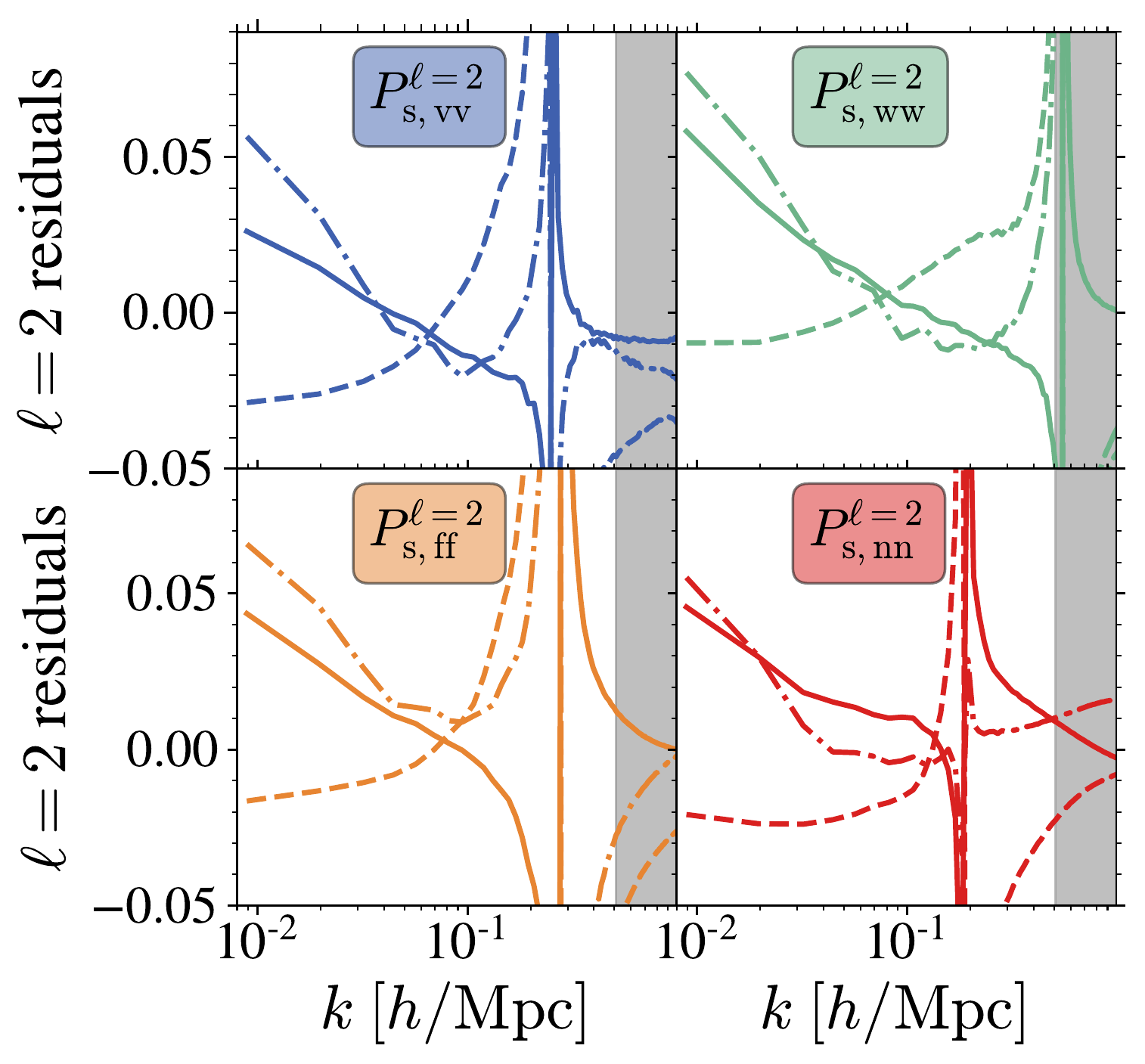}
    
    \caption{Residuals of the matter monopole (upper left panel) or quadrupole (upper right panel) and environment-dependent monopoles (lower left panel) or quadrupoles (lower right panel) when varying either $\sigma_\mathrm{8}$, $\Omega_\mathrm{m}$ or $M_\nu$. Residuals are defined as $P^{\ell}_\mathrm{s,\alpha\alpha}(k)^{\theta_i} / P^{\ell}_\mathrm{s,\alpha\alpha}(k)^{\mathrm{fid}} - 1$ with $\theta_i = M_\nu^+$ (solid line), $\sigma_\mathrm{8}^-$ (dashed line), or $\Omega_\mathrm{m}^-$ (dashed dotted line).}
    \label{fig:neutrinos_vs_sigma8}
\end{figure*}

In Fig.~\ref{fig:monopoles_real_rsd} are shown the ratios between the real-space and redshift-space monopoles obtained from the average of $7000$ fiducial simulations. As expected, all the spectra are getting boosted at large scales (small $k$ values) due to the coherent motion of matter escaping from voids and moving towards dense regions induced by the Kaiser effect. On the other hand, a decrease of power is observed at smaller scales where the FoG effect dominates, spreading particles initially residing in spherical overdensity along the line of sight. As expected, FoG mostly impacts the node environment where the highest velocities are found.
Beyond the shape of the spectra in redshift-space, the individual impact of each cosmological parameter is also different, as illustrated in Fig.~\ref{fig:neutrinos_vs_sigma8}. As an example, decreasing $\sigma_8$ does not imply a simple shift of the matter power spectrum monopole in redshift-space but is boosting the power at small scales (large $k$ values) due to the FoG effect. This is illustrated by the dashed lines of the top-left panel showing the monopole residuals $P^{\ell=0}_\mathrm{s,mm}(k)^{\theta_i} / P^{\ell=0}_\mathrm{s, mm}(k)^{\mathrm{fid}} - 1$ with $\theta_i$ being either $\sigma_\mathrm{8}^-$, $\Omega_\mathrm{m}^-$ or $M_\nu^+$. These effects can then cause degeneracies between cosmological parameters. It has been shown for instance that the effect of massive neutrinos can be mimicked by a decrease of $\sigma_\mathrm{8}$ on the redshift-space monopole at small scales $k>0.1$ $h$/Mpc \citep[see e.g.][]{Villaescusa-Navarro2018, Hahn2020, Bayer2021a}, which in turn has a similar impact as a decrease of $\Omega_\mathrm{m}$. Comparing to the bottom-left panels from Fig.~\ref{fig:neutrinos_vs_sigma8}, we see that monopoles computed in the cosmic web environments have various dependencies when decreasing the three parameters and inevitably, a change in $\sigma_8$ cannot be reproduced anymore by an increasing sum of neutrino mass, illustrating the breaking of degeneracies that we expect by analysing separately the several environment.
The anisotropic nature of the matter density field in redshift space however leads to nonzero $\ell>0$ multipoles that already have various shapes when varying these parameters, as illustrated in the upper right panel of Fig.~\ref{fig:neutrinos_vs_sigma8} for $\ell = 2$. This suggest that combining the two first poles of the matter density field should allow to break some degeneracies already. The measure of the multipoles are however noisier when going to larger $\ell$ and it might be of interest to see the reachable accuracy of the environment monopoles only.
We also see from the right-bottom panels that the environments themselves have different shapes for $\ell=2$, suggesting further possibilities for the statistics drawn from the environments to impose tighter constraints when combining the four monopoles and the four quadrupole derived in the several environments.

Obviously, the effects of the distortions are also rooted in the classification itself where we expect for instance the FoG leading to a leakage of some cells classified in nodes in real space to filaments in redshift-space.
These effects are quantified in Table~\ref{tab:confusion_matrix} reporting the confusion matrix averaged over ten simulations at fiducial cosmology when considering the real-space classification as the ground truth. Non-diagonal terms of this matrix hence correspond to the fraction of change in the environment classification when going from real to redshift space and a perfect match would lead to the identity matrix.
It shows for instance that $36.7$\% of the particles classified in nodes in real space are now found in filaments in redshift-space. We note that the leakage is mostly occurring between an environment and the one of directly higher local dimension (voids to walls, walls to filaments and filaments to nodes) while very few between other ones. We interpret this effect as a result of our plane-parallel approximation forcing the distortions to occur along one spatial axis only. It is worth underlining that, apart from nodes, the classification is quite stable to the distortions with about $80\%$ of the particles in each environment (voids, walls and filaments) keeping the same classification in real and redshift spaces.

\begin{table}
    \centering
    \caption{Averaged confusion matrix obtained from ten boxes of the Quijote simulations with the fiducial cosmology when considering the absence of RSD as a ground truth.}
    \label{tab:confusion_matrix}
    
    \smallskip
    
    \begin{tabular}{l|*{4}r}
    \toprule
    \diagbox[width=2.50cm, height=3.00cm]{\raisebox{5pt}{\hspace*{-0.10cm}Real space }}{\raisebox{0.57cm}{\rotatebox{90}{Redshift space}}} & \raisebox{-0.25cm}{\rotatebox{60}{\color{void}Void}} & \raisebox{-0.25cm}{\rotatebox{60}{\color{wall}Wall}} & \raisebox{-0.25cm}{\rotatebox{60}{\color{filament}Filament}} & \raisebox{-0.25cm}{\rotatebox{60}{\color{node}Node}} \\
    \midrule \\
    \hspace*{0.15cm}\color{void} Void & 0.813 & 0.179 & 0.007 & 0.001 \\ \\
    \hspace*{0.15cm}\color{wall} Wall & 0.018 & 0.809 & 0.166 & 0.006 \\ \\
    \hspace*{0.15cm}\color{filament} Filament & 0.001 & 0.082 & 0.809 & 0.107 \\ \\
    \hspace*{0.15cm}\color{node} Node & 0.002 & 0.042 & 0.367 & 0.587 \\ \\
    \bottomrule
    \end{tabular}
\end{table}

\begin{table*}
    \caption{Marginalised 1-$\sigma$ constraints obtained from the analysis of power spectra monopoles and quadrupoles computed in the different environments for all cosmological parameters. Improvement factors are relative to the matter case in redshift-space, namely $P^{\ell = \{0, 2\}}_\mathrm{s, mm}$ given in the third row. $\sigma_{M_\nu}$ is in unit of eV.}
    \centering
    \label{tab:constraints_RSD_with_quad}
    
    \renewcommand{\arraystretch}{1.5}
    \smallskip
    \begin{adjustbox}{max width=1.0\textwidth,center}
    \begin{tabular}{c|cccccc}
        \toprule
        Statistics & $\sigma_{\Omega_\mathrm{m}}$ & $\sigma_{\Omega_\mathrm{b}}$ & $\sigma_{h}$ & $\sigma_{n_\mathrm{s}}$ & $\sigma_{\sigma_\mathrm{8}}$ & $\sigma_{M_\nu}$ \\
        \midrule
        $P_\mathrm{mm}$ & $0.0969$ & $0.0413$ & $0.5145$ & $0.5019$ & $0.0132$ & $0.8749$ \\
        \midrule
        
        $P_\mathrm{s,mm}^{\ell=0}$ & $0.0081$ & $0.0144$ & $0.1495$ & $0.0786$ & $0.0115$ & $0.3611$ \\
        
        $P_\mathrm{s,mm}^{\ell=\{0,2\}}$ & $0.0046$ & $0.0133$ & $0.1396$ & $0.0719$ & $0.0020$ & $0.0834$ \\
        
        \midrule
        
        $P_\mathrm{s, comb}^{\ell=0}$ & $0.0033 \, (1.4)$ & $0.0105 \, (1.3)$ & $0.0877 \, (1.6)$ & $0.0346 \, (2.1)$ & $0.0026 \, (0.8)$ & $0.0482 \, (1.7)$ \\
        
        $P_\mathrm{s, comb}^{\ell=\{0,2\}}$ & $0.0027 \, (1.7)$ & $0.0097 \, (1.4)$ & $0.0773 \, (1.8)$ & $0.0295 \, (2.4)$ & $0.0020 \, (1)$ & $0.0304 \, (2.7)$ \\
        
        $P_\mathrm{s, comb}^{\ell=\{0,2\}} + P_\mathrm{s,mm}^{\ell=\{0,2\}}$ & $0.0011 \, (4.1)$ & $0.0091 \, (1.5)$ & $0.0716 \, (2.0)$ & $0.0279 \, (2.6)$ & $0.0015 \, (1.4)$ & $0.0151 \, (5.5)$ \\
        
        \bottomrule
    \end{tabular}
    \end{adjustbox}
    \end{table*}

    \begin{figure*}
        \centering
        \includegraphics[width=1.0\linewidth]{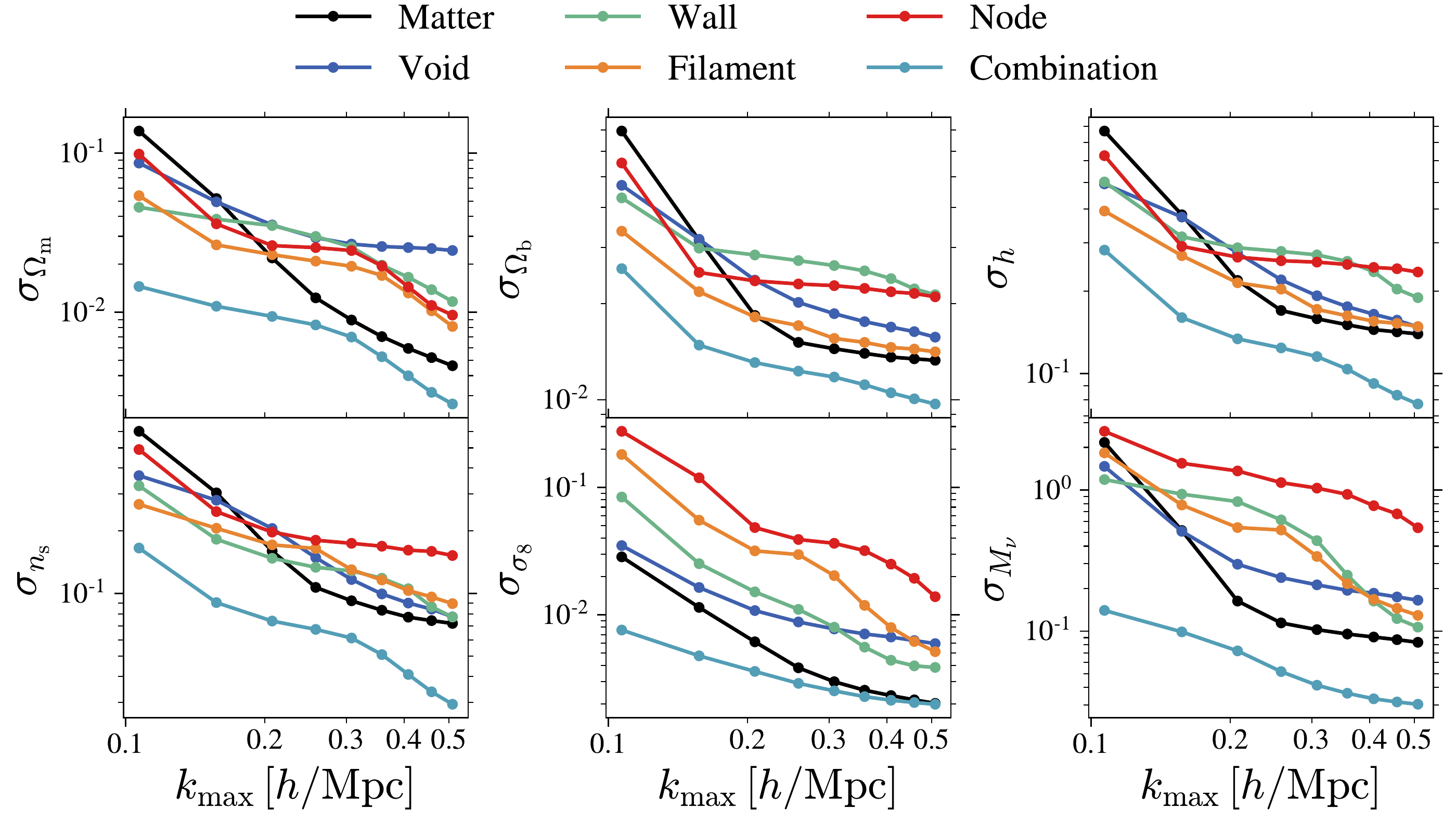}
        \caption{Evolution of the marginalised constraint $\sigma_{\theta_i}$ given by $P^{\ell=\{0, 2\}}_{\alpha\alpha}$ on cosmological parameters $\{\Omega_\mathrm{m}, \Omega_\mathrm{b}, h, n_\mathrm{s}, \sigma_8\}$ and the sum of neutrino mass $M_\nu$ with the maximum scale used for the Fisher analysis, $k_\mathrm{max}$. $\sigma_{M_\nu}$ is in unit of eV.}
        \label{fig:constraints_kmax}
    \end{figure*}
    
\begin{figure}
    \centering
    \includegraphics[width=1.0\linewidth]{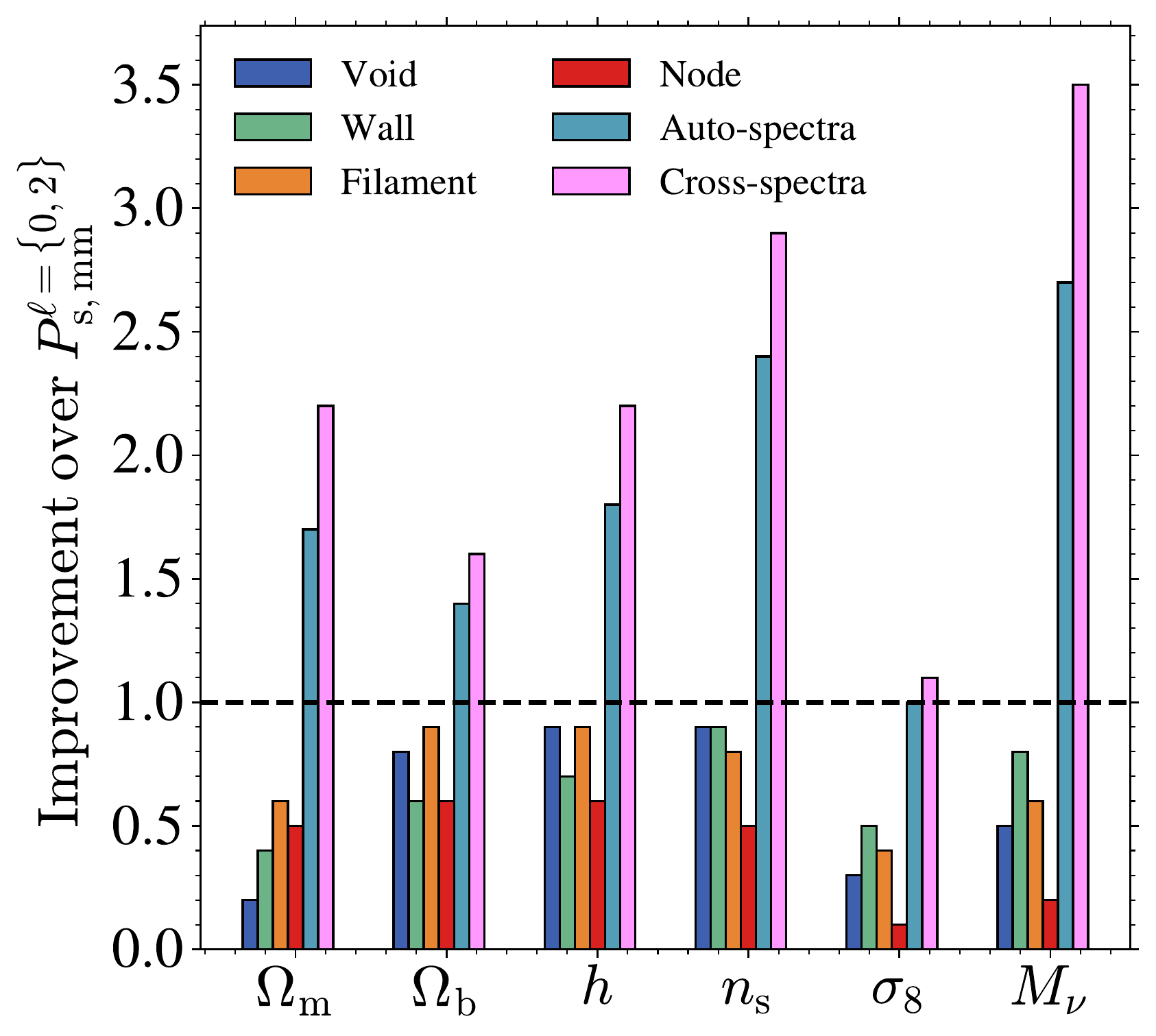}
    \caption{Improvement factors of the several studied statistics over the redshift-space matter monopole + quadrupole constraints for each of the six cosmological parameters at $k_{\mathrm{max}} = 0.5$ $h$/Mpc. The horizontal black line shows the unity improvement. Note that these statistics exclude the combination with the matter multipoles and concern uniquely the several cosmic web environments and their combination (therefore excluding the last line of Table~\ref{tab:constraints_RSD_with_quad} for instance).}
    \label{fig:barplot}
\end{figure}

\section{Cosmological constraints} \label{sect:constraints}

To quantify the amount of information carried by the spectra computed in the cosmic web environments, we make use of a Fisher analysis (details can be found in Appendix \ref{appendix:Fisher}).
In short, it allows us to derive a lower-bound on the constraints that one would reach when relying on a given statistic, in our case, either the spectra in the cosmic web environments or the full matter power spectrum. The derivation of the constraints mostly relies on two elements that we both numerically derive from the Quijote simulations: (i) the covariance matrix between Fourier amplitudes and (ii) the derivatives of the statistic with respect to the six cosmological parameters $\Omega_\mathrm{m}, \Omega_\mathrm{b}, h, n_\mathrm{s}, \sigma_8$ and $M_\nu$. These different components of the Fisher forecast are shown in Appendix \ref{appendix:cov_deriv}.

The first row of Table~\ref{tab:constraints_RSD_with_quad} reminds the constraints obtained in real space and presented in \cite{Bonnaire2021c} while the second and third rows contain the constraints from the matter monopole $P^{\ell = 0}_\mathrm{s, mm}$ and the combination of the monopole and quadrupole $P^{\ell = \{0, 2\}}_\mathrm{s, mm}$, respectively.
It is particularly striking that the redshift-space constraints obtained from just the monopole allows to improve by a factor of up to $12$ those in real space for $\Omega_{\mathrm{m}}$ or $2.4$ for $M_\nu$. These gains are even reaching $21.1$ and $10.5$ on the very same parameters when adding the information of the quadrupole in the $P_{\mathrm{s, mm}}^{\ell = \{0, 2\}}$ statistic. Such results are expected when moving from real to redshift space due to the additional velocity information coming into play in the breaking of the density field's isotropy.
It is also interesting to note that only $\sigma_{\sigma_8}$ and $\sigma_{M_\nu}$ are getting drastically reduced when adding the quadrupole information to matter monopole while errors on other parameters roughly stay still, quantifying the expected breaking of degeneracy between these two parameters discussed in Sect.~\ref{subsect:cw_redshift} and illustrated in the top panels of Fig.~\ref{fig:neutrinos_vs_sigma8}.

Concerning the environment-dependent spectra, we report in the lines four to six in Table~\ref{tab:constraints_RSD_with_quad} that:
\begin{itemize}
    \item Using the combination of environment monopoles (fourth line of Table~\ref{tab:constraints_RSD_with_quad}) already provides tighter constraints than the combination of the first two poles of the full matter density field, $P_\mathrm{s,mm}^{\ell=\{0,2\}}$ (third line of Table~\ref{tab:constraints_RSD_with_quad}) with up to $2.1$ and $1.7$ improvement factors for $n_\mathrm{s}$ and $M_\nu$.
    \item Adding the $\ell = 2$ information brings down the constraints to even lower values with improvement factors of $\{1.7, 1.4, 1.8, 2.4, 1, 2.7\}$ on cosmological parameters $\{\Omega_\mathrm{m}, \Omega_\mathrm{b}, h, n_\mathrm{s}, \sigma_8, M_\nu\}$ with respect to the matter analogue in redshift-space, as shown in the fifth line of Table~\ref{tab:constraints_RSD_with_quad}. Similarly to what is observed in the matter case, adding up the $\ell=2$ information mostly breaks the degeneracy on $M_\nu$ without impacting much the set of other parameters.
    \item The constraints imposed by the combination of environments statistics highly depends on the maximum scale of the spectra included in the analysis. Many panels of Fig.~\ref{fig:constraints_kmax} indeed indicate that the matter monopole+quadrupole saturates at mildly non-linear scales, a feature already exhibited in the real space analysis from \cite{Bonnaire2021c} and previously pointed out by \cite{Takahashi2010, Blot2015, Chan2017}. This saturation, mainly due to the degeneracies among parameters of the cosmological model, is not observed for the combination of environment-dependent spectra and especially encourages to push further the limiting scale of the analyses in order to fully benefit from the cosmic web split.
    \item Unlike the real-space case, the improvement factors of the individual environments are below one for all parameters when $k_{\mathrm{max}}=0.5$ $h$/Mpc (see Table~\ref{tab:constraints_envs}). This observation is not true at all scales, as illustrated by Fig.~\ref{fig:constraints_kmax} where they perform better than the matter statistics when restricting the analysis to very large scales, below $0.15$ $h$/Mpc in practice, expect for $\sigma_8$ and $M_\nu$ for which the full matter is always more attractive than individual environments.
    \item The absolute values of the constraints derived from individual environments and their combination in redshift space are similar to the real space ones (comparing Table~\ref{tab:constraints_envs} to Table C.1. in \cite{Bonnaire2021c}), and even tighter for $\Omega_{\mathrm{m}}$ and $M_{\nu}$.
    \item The non-negligible improvements of the constraints drawn by $P^{\ell=\{0, 2\}}_\mathrm{s, comb}$ are amplified when the analysis is reduced to large scales only (i.e. small values of $k$), especially for parameters like $\Omega_\mathrm{m}$, $\Omega_\mathrm{b}$, $\sigma_8$ and $M_\nu$ reaching up to a factor $5$ of improvement over the matter statistic, as represented in Fig.~\ref{fig:constraints_kmax}.
    \item The overall information brought by the combination of environment-dependant spectra can still be completed by the matter multipoles to further improve the constraint by factors up to $5.5$ and $4.1$ on $M_{\nu}$ and $\Omega_{\mathrm{m}}$ respectively (see last row of Table~\ref{tab:constraints_RSD_with_quad}). It seems indeed, by inspection of the confidence ellipses in Fig.~\ref{fig:corner}, that $P_{\mathrm{s, comb}}^{\ell=\{0, 2\}}$ and $P_{\mathrm{s, mm}}^{\ell=\{0, 2\}}$ bring complementary information in panels like $M_\nu$--$\sigma_8$ or $M_\nu$--$\Omega_{\mathrm{m}}$ explaining the tighter constraints obtained when merging these information in the analysis.
    \item Including the cross-spectra in the combination, hence considering the statistic $P_{\mathrm{s, \alpha\beta}}^{\ell=\{0, 2\}}$ for $\left(\alpha,\beta\right) \in \{\mathrm{v}, \mathrm{w}, \mathrm{f}, \mathrm{n}\}^2$ allows to boost the gains on some parameters like $M_\nu$ and $\Omega_{\mathrm{m}}$ to $2.2$ and $3.5$ respectively. All these gains are summarised in Fig.~\ref{fig:barplot} and this statistic involves notably the information from auto-spectra and accounts for a total of $800$ elements in the summary statistic 
\end{itemize}

We have also checked that including the hexadecapole ($\ell=4$) in our analysis does not lead to any further improvement of the constraints over the $\ell=\{0, 2\}$ statistics, neither for the matter, nor for the environment-dependant spectra and their combination.

\section{Discussion and conclusion}  \label{sect:disc_ccl}

The analysis of the matter distribution in redshift space, summarised by its power spectrum, is know to greatly improve the cosmological constraints thanks to the velocity information contained in the anisotropy of the density fields which breaks key degeneracies between parameters.
Our results show that we can attain even tighter constraints by performing the statistical analysis in identified cosmic-web environments. In this configuration, the constraints on cosmological parameters such as the initial spectral index, the matter density, the Hubble parameter and most notably the summed neutrino mass can be improved by up to factors two, and even reach five when combining with the full matter statistics. Our approach hence opens up a new avenue of analysing the observed distribution of tracers to derive cosmological constraints. It especially shows that conditioning the correlations on the cosmic web environments improves the constraints on the cosmological parameters over a direct correlation of all tracers, in agreement with previous findings \citep[e.g.,][]{Uhlemann2020, Paillas2021, Paillas2022}.

The present study is performed in a theoretical context in which we have access to the matter density field from numerical simulations. 
Firstly, within this idealised context and with the studied statistics, it appears that constraining the cosmological parameters in redshift space yields similar -- if not tighter -- constraints than in real space, even though we do not handle the distortions in the classification. This therefore suggests that reconstruction of the particles from redshift to real space \cite[e.g.][]{Jasche2013, Bos2014, Leclercq2015} is not necessary in the present setting and can even deteriorates the accuracy of the derived constraints.
Secondly, this context allows us to explore the full range scales including the most non-linear ones. We show that the combination of auto-spectra computed in the different environments up to non-linear scales, $k_\mathrm{max} = 0.5$ $h$/Mpc, is in favour of redshift-space statistics for most parameters, even more emphasised when adding information from the full matter monopole and quadrupole.

With this second paper of the series, we hence go beyond showing how the use of the cosmic web environments as a leverage to improve the cosmological constraints in real space and present the benefits of a cosmic web classification to improve redshift-space constraints over the traditional matter power spectrum multipoles. In a third paper of the series, we will complete our comprehensive analysis of the cosmological content of the cosmic web environments with the analysis of the direct higher order statistic, namely the bispectrum.
The next steps and challenges will consist in adapting and optimising our analysis, currently performed in a theoretical and idealised setup of large number density provided by particles in simulations, to more realistic cases of observed sparse tracers of the density field, like halos or galaxies, as performed in other works \citep[such as][]{Hahn2020, Naidoo2021, Kreisch2021, Paillas2022}. In this context, simulations will represent a major endeavour. Adapting our analysis so that it can be applied to actual data from galaxy survey require to be tested against simulations with a higher mass resolution enabling the definition of galaxy-type halos and with realistic baryonic physics models \cite{Vogelsberger2014}. 
Such simulations will also allow to bypass the difficulty of a theoretical modelling of covariance matrices and building of likelihoods, via the use of simulation-based inference \cite[e.g.,][]{Cranmer2020b, Lemos2022}. The promises of the cosmological content of cosmic web environments exhibited by our results will surely be worth the efforts by constituting a step toward a optimal and interpretable summary statistics that will allows to take full advantage of the future Stage IV experiments \citep{Leureijs2011, WFIRST15, DESICollaboration2016}.

\begin{acknowledgements}
The authors thank the members of the ByoPiC team\footnote{\url{https://byopic.eu/team}} for useful discussions.
They also thank the Quijote team for making their data publicly available and particularly F. Villaescusa-Navarro for his help and availability regarding the Quijote suite.
This research was supported by funding for the ByoPiC project from the European Research Council (ERC) under the European Union’s Horizon 2020 research and innovation program grant number ERC-2015-AdG 695561.
A.D. was supported by the Comunidad de Madrid and the Complutense University of Madrid (Spain) through the Atracción de Talento program (Ref. 2019-T1/TIC-13298).
T.B. acknowledges funding from the French government under management of Agence Nationale de la Recherche as part of the ``Investissements d’avenir'' program, reference ANR-19-P3IA-0001 (PRAIRIE 3IA Institute).
\end{acknowledgements}

\medskip

\bibliographystyle{aa}
\bibliography{thesis}

\begin{thebibliography}{71}
\expandafter\ifx\csname natexlab\endcsname\relax\def\natexlab#1{#1}\fi

\bibitem[{Abbott {et~al.}(2019)Abbott, Allam, Andersen, Angus, Asorey, Avelino,
  Avila, Bassett, Bechtol, Bernstein, Bertin, Brooks, Brout, Brown, Burke,
  Calcino, Rosell, Carollo, {Carrasco Kind}, Carretero, Casas, Castander,
  Cawthon, Challis, Childress, Clocchiatti, Cunha, D'Andrea, da~Costa, Davis,
  Davis, {De Vicente}, DePoy, Desai, Diehl, Doel, Drlica-Wagner, Eifler,
  Evrard, Fernandez, Filippenko, Finley, Flaugher, Foley, Fosalba, Frieman,
  Galbany, Garc{\'{i}}a-Bellido, Gaztanaga, Giannantonio, Glazebrook,
  Goldstein, Gonz{\'{a}}lez-Gait{\'{a}}n, Gruen, Gruendl, Gschwend, Gupta,
  Gutierrez, Hartley, Hinton, Hollowood, Honscheid, Hoormann, Hoyle, James,
  Jeltema, Johnson, Johnson, Kasai, Kent, Kessler, Kim, Kirshner, Kovacs,
  Krause, Kron, Kuehn, Kuhlmann, Kuropatkin, Lahav, Lasker, Lewis, Li, Lidman,
  Lima, Lin, Macaulay, Maia, Mandel, March, Marriner, Marshall, Martini,
  Menanteau, Miller, Miquel, Miranda, Mohr, Morganson, Muthukrishna,
  M{\"{o}}ller, Neilsen, Nichol, Nord, Nugent, Ogando, Palmese, Pan, Plazas,
  Pursiainen, Romer, Roodman, Rozo, Rykoff, Sako, Sanchez, Scarpine, Schindler,
  Schubnell, Scolnic, Serrano, Sevilla-Noarbe, Sharp, Smith, Soares-Santos,
  Sobreira, Sommer, Spinka, Suchyta, Sullivan, Swann, Tarle, Thomas, Thomas,
  Troxel, Tucker, Uddin, Walker, Wester, Wiseman, Wolf, Yanny, Zhang, \&
  Zhang}]{Abbott2018}
Abbott, T. M.~C., Allam, S., Andersen, P., {et~al.} 2019, The Astrophysical
  Journal, 872, L30

\bibitem[{Alam {et~al.}(2020)Alam, Arnold, Aviles, Bean, Cai, Cautun,
  Cervantes-Cota, Cuesta-Lazaro, Devi, Eggemeier, Fromenteau, Gonzalez-Morales,
  Halenka, He, Hellwing, Hernandez-Aguayo, Ishak, Koyama, Li, de~la Macorra,
  Rizo, Miller, Mueller, Niz, Ntelis, Otero, Sabiu, Slepian, Stark, Valenzuela,
  Valogiannis, Vargas-Magana, Winther, Zarrouk, Zhao, \& Zheng}]{Alam2020}
Alam, S., Arnold, C., Aviles, A., {et~al.} 2020, arXiv e-prints, 1

\bibitem[{Alam {et~al.}(2017{\natexlab{a}})Alam, Ata, Bailey, Beutler, Bizyaev,
  Jonathan, Bolton, Brownstein, Burden, Chuang, Cuesta, Dawson, Eisenstein,
  Escoffier, Kitaura, Malanushenko, Malanushenko, Maraston, Mcbride, Nichol,
  Olmstead, Oravetz, Palanque-delabrouille, Pan, Pellejero-ibanez, Will,
  Petitjean, Prada, Price-whelan, Reid, Rodr, Roe, Ross, Ross, Rossi, Rubi,
  Samushia, Satpathy, Sc, Schlegel, Donald, Seo, Simmons, Thomas, Tinker,
  Tojeiro, Maga, Al, Verde, Wake, Wang, Weinberg, White, Wood-vasey, \&
  Christophe}]{Alam2016}
Alam, S., Ata, M., Bailey, S., {et~al.} 2017{\natexlab{a}}, Monthly Notices of
  the Royal Astronomical Society, 470, 2617

\bibitem[{Alam {et~al.}(2017{\natexlab{b}})Alam, Miyatake, More, Ho, \&
  Mandelbaum}]{Alam2017}
Alam, S., Miyatake, H., More, S., Ho, S., \& Mandelbaum, R. 2017{\natexlab{b}},
  Monthly Notices of the Royal Astronomical Society, 465, 4853

\bibitem[{Allys {et~al.}(2020)Allys, Marchand, Cardoso, Villaescusa-Navarro,
  Ho, \& Mallat}]{Allys2020}
Allys, E., Marchand, T., Cardoso, J.~F., {et~al.} 2020, Physical Review D, 102,
  103506

\bibitem[{Aragon-Calvo {et~al.}(2010)Aragon-Calvo, Weygaert, \&
  Jones}]{AragonCalvo2010}
Aragon-Calvo, M., Weygaert, R. V.~D., \& Jones, B. J.~T. 2010, Monthly Notices
  of the Royal Astronomical Society, 408, 2163

\bibitem[{Bayer {et~al.}(2022)Bayer, Banerjee, \& Seljak}]{Bayer2021a}
Bayer, A.~E., Banerjee, A., \& Seljak, U. 2022, Physical Review D, 105

\bibitem[{Bayer {et~al.}(2021)Bayer, Villaescusa-Navarro, Massara, Liu,
  Spergel, Verde, Wandelt, Viel, \& Ho}]{Bayer2021}
Bayer, A.~E., Villaescusa-Navarro, F., Massara, E., {et~al.} 2021, The
  Astrophysical Journal, 919, 24

\bibitem[{Blake {et~al.}(2020)Blake, Amon, Asgari, Bilicki, Dvornik, Erben,
  Giblin, Glazebrook, Heymans, Hildebrandt, Joachimi, Joudaki, Kannawadi,
  Kuijken, Lidman, Parkinson, Shan, Tr{\"{o}}ster, {Van Den Busch}, Wolf, \&
  Wright}]{Blake2020}
Blake, C., Amon, A., Asgari, M., {et~al.} 2020, Astronomy and Astrophysics,
  642, 1

\bibitem[{Blot {et~al.}(2015)Blot, Corasaniti, Alimi, Reverdy, \&
  Rasera}]{Blot2015}
Blot, L., Corasaniti, P.~S., Alimi, J.~M., Reverdy, V., \& Rasera, Y. 2015,
  Monthly Notices of the Royal Astronomical Society, 446, 1756

\bibitem[{Bonnaire {et~al.}(2020)Bonnaire, Aghanim, Decelle, \&
  Douspis}]{Bonnaire2020}
Bonnaire, T., Aghanim, N., Decelle, A., \& Douspis, M. 2020, Astronomy and
  Astrophysics, 637, A18

\bibitem[{Bonnaire {et~al.}(2021)Bonnaire, Aghanim, Kuruvilla, \&
  Decelle}]{Bonnaire2021c}
Bonnaire, T., Aghanim, N., Kuruvilla, J., \& Decelle, A. 2021, Astronomy {\&}
  Astrophysics, 661, A146

\bibitem[{Bonnaire {et~al.}(2022)Bonnaire, Decelle, \& Aghanim}]{Bonnaire2021b}
Bonnaire, T., Decelle, A., \& Aghanim, N. 2022, IEEE Transactions on Pattern
  Analysis and Machine Intelligence, 44, 9119

\bibitem[{Bos {et~al.}(2014)Bos, {Van De Weygaert}, Kitaura, \&
  Cautun}]{Bos2014}
Bos, E.~G., {Van De Weygaert}, R., Kitaura, F., \& Cautun, M. 2014, Proceedings
  of the International Astronomical Union, 11, 271

\bibitem[{Carron(2013)}]{Carron2013}
Carron, J. 2013, Astronomy and Astrophysics, 551, 10

\bibitem[{Cautun {et~al.}(2013)Cautun, van~de Weygaert, \& Jones}]{Nexus}
Cautun, M., van~de Weygaert, R., \& Jones, B.~J. 2013, Monthly Notices of the
  Royal Astronomical Society, 429, 1286

\bibitem[{Chan \& Blot(2017)}]{Chan2017}
Chan, K.~C. \& Blot, L. 2017, Physical Review D, 96, 023528

\bibitem[{Cheng \& M{\'{e}}nard(2021)}]{Cheng2021}
Cheng, S. \& M{\'{e}}nard, B. 2021, Monthly Notices of the Royal Astronomical
  Society, 8, 1

\bibitem[{Cheng {et~al.}(2020)Cheng, Ting, M{\'{e}}nard, \& Bruna}]{Cheng2020}
Cheng, S., Ting, Y.-S., M{\'{e}}nard, B., \& Bruna, J. 2020, Monthly Notices of
  the Royal Astronomical Society, 499, 5902

\bibitem[{Cranmer {et~al.}(2020)Cranmer, Brehmer, \& Louppe}]{Cranmer2020b}
Cranmer, K., Brehmer, J., \& Louppe, G. 2020, Proceedings of the National
  Academy of Sciences, 117, 30055

\bibitem[{{DESI Collaboration} {et~al.}(2016){DESI Collaboration}, Aghamousa,
  Aguilar, Ahlen, Alam, Allen, Prieto, Annis, Bailey, Balland, Ballester,
  Baltay, Beaufore, Bebek, Beers, Bell, Bernal, Besuner, Beutler, Blake,
  Bleuler, Blomqvist, Blum, Bolton, Briceno, Brooks, Brownstein, Buckley-Geer,
  Burden, Burtin, Busca, Cahn, Cai, Cardiel-Sas, Carlberg, Carton, Casas,
  Castander, Cervantes-Cota, Claybaugh, Close, Coker, Cole, Comparat, Cooper,
  Cousinou, Crocce, Cuby, Cunningham, Davis, Dawson, de~la Macorra, {De
  Vicente}, Delubac, Derwent, Dey, Dhungana, Ding, Doel, Duan, Ealet,
  Edelstein, Eftekharzadeh, Eisenstein, Elliott, Escoffier, Evatt, Fagrelius,
  Fan, Fanning, Farahi, Farihi, Favole, Feng, Fernandez, Findlay, Finkbeiner,
  Fitzpatrick, Flaugher, Flender, Font-Ribera, Forero-Romero, Fosalba, Frenk,
  Fumagalli, Gaensicke, Gallo, Garcia-Bellido, Gaztanaga, Fusillo, Gerard,
  Gershkovich, Giannantonio, Gillet, Gonzalez-de Rivera, Gonzalez-Perez, Gott,
  Graur, Gutierrez, Guy, Habib, Heetderks, Heetderks, Heitmann, Hellwing,
  Herrera, Ho, Holland, Honscheid, Huff, Hutchinson, Huterer, Hwang, Laguna,
  Ishikawa, Jacobs, Jeffrey, Jelinsky, Jennings, Jiang, Jimenez, Johnson,
  Joyce, Jullo, Juneau, Kama, Karcher, Karkar, Kehoe, Kennamer, Kent,
  Kilbinger, Kim, Kirkby, Kisner, Kitanidis, Kneib, Koposov, Kovacs, Koyama,
  Kremin, Kron, Kronig, Kueter-Young, Lacey, Lafever, Lahav, Lambert, Lampton,
  Landriau, Lang, Lauer, Goff, Guillou, {Van Suu}, Lee, Lee, Leitner, Lesser,
  Levi, L'Huillier, Li, Liang, Lin, Linder, Loebman, Luki{\'{c}}, Ma, MacCrann,
  Magneville, Makarem, Manera, Manser, Marshall, Martini, Massey, Matheson,
  McCauley, McDonald, McGreer, Meisner, Metcalfe, Miller, Miquel, Moustakas,
  Myers, Naik, Newman, Nichol, Nicola, da~Costa, Nie, Niz, Norberg, Nord,
  Norman, Nugent, O'Brien, Oh, Olsen, Padilla, Padmanabhan, Padmanabhan,
  Palanque-Delabrouille, Palmese, Pappalardo, P{\^{a}}ris, Park, Patej,
  Peacock, Peiris, Peng, Percival, Perruchot, Pieri, Pogge, Pollack, Poppett,
  Prada, Prakash, Probst, Rabinowitz, Raichoor, Ree, Refregier, Regal, Reid,
  Reil, Rezaie, Rockosi, Roe, Ronayette, Roodman, Ross, Ross, Rossi, Rozo,
  Ruhlmann-Kleider, Rykoff, Sabiu, Samushia, Sanchez, Sanchez, Schlegel,
  Schneider, Schubnell, Secroun, Seljak, Seo, Serrano, Shafieloo, Shan,
  Sharples, Sholl, Shourt, Silber, Silva, Sirk, Slosar, Smith, Smoot, Som,
  Song, Sprayberry, Staten, Stefanik, Tarle, Tie, Tinker, Tojeiro, Valdes,
  Valenzuela, Valluri, Vargas-Magana, Verde, Walker, Wang, Wang, Weaver,
  Weaverdyck, Wechsler, Weinberg, White, Yang, Yeche, Zhang, Zhao, Zheng, Zhou,
  Zhou, Zhu, Zou, \& Zu}]{DESICollaboration2016}
{DESI Collaboration}, Aghamousa, A., Aguilar, J., {et~al.} 2016, arXiv e-prints
  [\eprint[arXiv]{1611.00036}]

\bibitem[{Drlica-Wagner {et~al.}(2019)Drlica-Wagner, Mao, Adhikari, Armstrong,
  Banerjee, Banik, Bechtol, Bird, Boddy, Bonaca, Bovy, Buckley, Bulbul, Chang,
  Chapline, Cohen-Tanugi, Cuoco, Cyr-Racine, Dawson, Rivero, Dvorkin, Erkal,
  Fassnacht, Garc{\'{i}}a-Bellido, Giannotti, Gluscevic, Golovich, Hendel,
  Hezaveh, Horiuchi, Jee, Kaplinghat, Keeton, Koposov, Lam, Li, Lu, Mandelbaum,
  McDermott, McNanna, Medford, Meyer, Marc, Murgia, Nadler, Necib, Nuss, Pace,
  Peter, Polin, Prescod-Weinstein, Read, Rosenfeld, Shipp, Simon, Slatyer,
  Straniero, Strigari, Tollerud, Tyson, Wang, Wechsler, Wittman, Yu, Zaharijas,
  Ali-Ha{\"{i}}moud, Annis, Birrer, Biswas, Blazek, Brooks, Buckley-Geer,
  Caputo, Charles, Digel, Dodelson, Flaugher, Frieman, Gawiser, Hearin,
  Hlo{\v{z}}ek, Jain, Jeltema, Koushiappas, Lisanti, LoVerde, Mishra-Sharma,
  Newman, Nord, Nourbakhsh, Ritz, Robertson, S{\'{a}}nchez-Conde, Slosar, Tait,
  Verma, Vilalta, Walter, Yanny, \& Zentner}]{Drlica-Wagner2019}
Drlica-Wagner, A., Mao, Y.-Y., Adhikari, S., {et~al.} 2019, arXiv e-prints
  [\eprint[arXiv]{1902.01055}]

\bibitem[{Eickenberg {et~al.}(2022)Eickenberg, Allys, Dizgah, Lemos, Massara,
  Abidi, Hahn, Hassan, Blancard, Ho, Mallat, And{\'{e}}n, \&
  Villaescusa-Navarro}]{Eickenberg2022}
Eickenberg, M., Allys, E., Dizgah, A.~M., {et~al.} 2022, arXiv e-prints
  [\eprint[arXiv]{2204.07646}]

\bibitem[{Eisenstein {et~al.}(2005)Eisenstein, Zehavi, Hogg, Scoccimarro,
  Blanton, Nichol, Scranton, Seo, Tegmark, Zheng, Anderson, Annis, Bahcall,
  Brinkmann, Burles, Castander, Connolly, Csabai, Doi, Fukugita, Frieman,
  Glazebrook, Gunn, Hendry, Hennessy, Ivezi{\'{c}}, Kent, Knapp, Lin, Loh,
  Lupton, Margon, McKay, Meiksin, Munn, Pope, Richmond, Schlegel, Schneider,
  Shimasaku, Stoughton, Strauss, SubbaRao, Szalay, Szapudi, Tucker, Yanny, \&
  York}]{Eisenstein2005}
Eisenstein, D.~J., Zehavi, I., Hogg, D.~W., {et~al.} 2005, The Astrophysical
  Journal, 633, 560

\bibitem[{Forero-Romero {et~al.}(2009)Forero-Romero, Hoffman, Gottl{\"{o}}ber,
  Klypin, \& Yepes}]{Forero-Romero2009}
Forero-Romero, J.~E., Hoffman, Y., Gottl{\"{o}}ber, S., Klypin, A., \& Yepes,
  G. 2009, Monthly Notices of the Royal Astronomical Society, 396, 1815

\bibitem[{Gualdi {et~al.}(2021)Gualdi, Gil-mar{\'{i}}n, \& Verde}]{Gualdi2021}
Gualdi, D., Gil-mar{\'{i}}n, H., \& Verde, L. 2021, Journal of Cosmology and
  Astroparticle Physics, 2021, 008

\bibitem[{Hahn \& Villaescusa-Navarro(2021)}]{Hahn2021}
Hahn, C.~H. \& Villaescusa-Navarro, F. 2021, Journal of Cosmology and
  Astroparticle Physics, 2021, 029

\bibitem[{Hahn {et~al.}(2020)Hahn, Villaescusa-Navarro, Castorina, \&
  Scoccimarro}]{Hahn2020}
Hahn, C.~H., Villaescusa-Navarro, F., Castorina, E., \& Scoccimarro, R. 2020,
  Journal of Cosmology and Astroparticle Physics, 2020, 0

\bibitem[{Hahn {et~al.}(2007)Hahn, Porciani, Carollo, \& Dekel}]{Hahn2007}
Hahn, O., Porciani, C., Carollo, C.~M., \& Dekel, A. 2007, Monthly Notices of
  the Royal Astronomical Society, 375, 489

\bibitem[{Hartlap {et~al.}(2007)Hartlap, Simon, \& Schneider}]{Hartlap2007}
Hartlap, J., Simon, P., \& Schneider, P. 2007, Astronomy and Astrophysics, 464,
  399

\bibitem[{Hildebrandt {et~al.}(2017)Hildebrandt, Viola, Heymans, Joudaki,
  Kuijken, Blake, Erben, Joachimi, Klaes, Miller, Morrison, Nakajima, {Verdoes
  Kleijn}, Amon, Choi, Covone, de~Jong, Dvornik, {Fenech Conti}, Grado,
  Harnois-D{\'{e}}raps, Herbonnet, Hoekstra, K{\"{o}}hlinger, McFarland, Mead,
  Merten, Napolitano, Peacock, Radovich, Schneider, Simon, Valentijn, van~den
  Busch, van Uitert, \& {Van Waerbeke}}]{Hildebrandt2017}
Hildebrandt, H., Viola, M., Heymans, C., {et~al.} 2017, Monthly Notices of the
  Royal Astronomical Society, 465, 1454

\bibitem[{Ivanov {et~al.}(2020)Ivanov, Simonovi{\'{c}}, \&
  Zaldarriaga}]{Ivanov2020}
Ivanov, M.~M., Simonovi{\'{c}}, M., \& Zaldarriaga, M. 2020, Journal of
  Cosmology and Astroparticle Physics, 2020 [\eprint[arXiv]{1909.05277}]

\bibitem[{Jackson(1972)}]{Jackson1972}
Jackson, J. 1972, Monthly Notices of the Royal Astronomical Society, 156, 74

\bibitem[{Jasche \& Wandelt(2013)}]{Jasche2013}
Jasche, J. \& Wandelt, B.~D. 2013, Monthly Notices of the Royal Astronomical
  Society, 432, 894

\bibitem[{Jullo {et~al.}(2019)Jullo, {De La Torre}, Cousinou, Escoffier,
  Giocoli, Metcalf, Comparat, Shan, Makler, Kneib, Prada, Yepes, \&
  Gottl{\"{o}}ber}]{Jullo2019}
Jullo, E., {De La Torre}, S., Cousinou, M.~C., {et~al.} 2019, Astronomy and
  Astrophysics, 627 [\eprint[arXiv]{1903.07160}]

\bibitem[{Kaiser(1987)}]{Kaiser1987}
Kaiser, N. 1987, Monthly Notices of the Royal Astronomical Society, 227, 1

\bibitem[{Kaufman(1967)}]{Kaufman1967}
Kaufman, G. 1967, Center for Operations Research and Econometrics, 44

\bibitem[{Kodwani {et~al.}(2019)Kodwani, Alonso, \& Ferreira}]{Kodwani2019}
Kodwani, D., Alonso, D., \& Ferreira, P.~G. 2019, The Open Journal of
  Astrophysics, 2, 3

\bibitem[{Kreisch {et~al.}(2022)Kreisch, Pisani, Villaescusa-Navarro, Spergel,
  Wandelt, Hamaus, \& Bayer}]{Kreisch2021}
Kreisch, C.~D., Pisani, A., Villaescusa-Navarro, F., {et~al.} 2022, The
  Astrophysical Journal, 935, 100

\bibitem[{Kuruvilla(2022)}]{Kuruvilla2021a}
Kuruvilla, J. 2022, Astronomy and Astrophysics, 660

\bibitem[{Kuruvilla \& Aghanim(2021)}]{Kuruvilla2021}
Kuruvilla, J. \& Aghanim, N. 2021, Astronomy and Astrophysics, 653, A130

\bibitem[{Laureijs {et~al.}(2011)Laureijs, Amiaux, Arduini, Augu{\`{e}}res,
  Brinchmann, Cole, Cropper, Dabin, Duvet, Ealet, Garilli, Gondoin, Guzzo,
  Hoar, Hoekstra, Holmes, Kitching, Maciaszek, Mellier, Pasian, Percival,
  Rhodes, Criado, Sauvage, Scaramella, Valenziano, Warren, Bender, Castander,
  Cimatti, F{\`{e}}vre, Kurki-Suonio, Levi, Lilje, Meylan, Nichol, Pedersen,
  Popa, Lopez, Rix, Rottgering, Zeilinger, Grupp, Hudelot, Massey, Meneghetti,
  Miller, Paltani, Paulin-Henriksson, Pires, Saxton, Schrabback, Seidel, Walsh,
  Aghanim, Amendola, Bartlett, Baccigalupi, Beaulieu, Benabed, Cuby, Elbaz,
  Fosalba, Gavazzi, Helmi, Hook, Irwin, Kneib, Kunz, Mannucci, Moscardini, Tao,
  Teyssier, Weller, Zamorani, Osorio, Boulade, Foumond, {Di Giorgio},
  Guttridge, James, Kemp, Martignac, Spencer, Walton, Bl{\"{u}}mchen, Bonoli,
  Bortoletto, Cerna, Corcione, Fabron, Jahnke, Ligori, Madrid, Martin,
  Morgante, Pamplona, Prieto, Riva, Toledo, Trifoglio, Zerbi, Abdalla, Douspis,
  Grenet, Borgani, Bouwens, Courbin, Delouis, Dubath, Fontana, Frailis,
  Grazian, Koppenh{\"{o}}fer, Mansutti, Melchior, Mignoli, Mohr, Neissner,
  Noddle, Poncet, Scodeggio, Serrano, Shane, Starck, Surace, Taylor,
  Verdoes-Kleijn, Vuerli, Williams, Zacchei, Altieri, Sanz, Kohley,
  Oosterbroek, Astier, Bacon, Bardelli, Baugh, Bellagamba, Benoist, Bianchi,
  Biviano, Branchini, Carbone, Cardone, Clements, Colombi, Conselice, Cresci,
  Deacon, Dunlop, Fedeli, Fontanot, Franzetti, Giocoli, Garcia-Bellido, Gow,
  Heavens, Hewett, Heymans, Holland, Huang, Ilbert, Joachimi, Jennins, Kerins,
  Kiessling, Kirk, Kotak, Krause, Lahav, van Leeuwen, Lesgourgues, Lombardi,
  Magliocchetti, Maguire, Majerotto, Maoli, Marulli, Maurogordato, McCracken,
  McLure, Melchiorri, Merson, Moresco, Nonino, Norberg, Peacock, Pello, Penny,
  Pettorino, {Di Porto}, Pozzetti, Quercellini, Radovich, Rassat, Roche,
  Ronayette, Rossetti, Sartoris, Schneider, Semboloni, Serjeant, Simpson,
  Skordis, Smadja, Smartt, Spano, Spiro, Sullivan, Tilquin, Trotta, Verde,
  Wang, Williger, Zhao, Zoubian, \& Zucca}]{Leureijs2011}
Laureijs, R., Amiaux, J., Arduini, S., {et~al.} 2011, arXiv e-prints
  [\eprint[arXiv]{1110.3193}]

\bibitem[{Leclercq {et~al.}(2015)Leclercq, Jasche, Lavaux, \&
  Wandelt}]{Leclercq2015}
Leclercq, F., Jasche, J., Lavaux, G., \& Wandelt, B. 2015, arXiv e-prints
  [\eprint[arXiv]{1512.02242}]

\bibitem[{Lemos {et~al.}(2022)Lemos, Cranmer, Abidi, Hahn, Eickenberg, Massara,
  Yallup, \& Ho}]{Lemos2022}
Lemos, P., Cranmer, M., Abidi, M., {et~al.} 2022 [\eprint[arXiv]{2207.08435}]

\bibitem[{Libeskind {et~al.}(2017)Libeskind, van~de Weygaert, Cautun, Falck,
  Tempel, Abel, Alpaslan, Arag{\'{o}}n-Calvo, Forero-Romero, Gonzalez,
  Gottl{\"{o}}ber, Hahn, Hellwing, Hoffman, Jones, Kitaura, Knebe, Manti,
  Neyrinck, Nuza, Padilla, Platen, Ramachandra, Robotham, Saar, Shandarin,
  Steinmetz, Stoica, Sousbie, \& Yepes}]{Libeskind2017}
Libeskind, N.~I., van~de Weygaert, R., Cautun, M., {et~al.} 2017, Monthly
  Notices of the Royal Astronomical Society, 473, 1195

\bibitem[{Mandelbaum {et~al.}(2013)Mandelbaum, Slosar, Baldauf, Seljak, Hirata,
  Nakajima, Reyes, \& Smith}]{Mandelbaum2013}
Mandelbaum, R., Slosar, A., Baldauf, T., {et~al.} 2013, Monthly Notices of the
  Royal Astronomical Society, 432, 1544

\bibitem[{Martizzi {et~al.}(2019)Martizzi, Vogelsberger, Artale, Haider,
  Torrey, Marinacci, Nelson, Pillepich, Weinberger, Hernquist, Naiman, \&
  Springel}]{Martizzi2018}
Martizzi, D., Vogelsberger, M., Artale, M.~C., {et~al.} 2019, Monthly Notices
  of the Royal Astronomical Society, 486, 3766

\bibitem[{Massara {et~al.}(2022)Massara, Villaescusa-Navarro, Hahn, Abidi,
  Eickenberg, Ho, Lemos, Dizgah, \& Blancard}]{Massara2022}
Massara, E., Villaescusa-Navarro, F., Hahn, C., {et~al.} 2022, arXiv e-prints
  [\eprint[arXiv]{2206.01709}]

\bibitem[{Massara {et~al.}(2021)Massara, Villaescusa-Navarro, Ho, Dalal, \&
  Spergel}]{Massara2021}
Massara, E., Villaescusa-Navarro, F., Ho, S., Dalal, N., \& Spergel, D.~N.
  2021, Physical Review Letters, 126, 1

\bibitem[{Mueller {et~al.}(2015)Mueller, Bernardis, Bean, \&
  Niemack}]{Mueller2015}
Mueller, E.~M., Bernardis, F.~D., Bean, R., \& Niemack, M.~D. 2015,
  Astrophysical Journal, 808, 47

\bibitem[{Naidoo {et~al.}(2021)Naidoo, Massara, \& Lahav}]{Naidoo2021}
Naidoo, K., Massara, E., \& Lahav, O. 2021, arXiv e-prints, 1

\bibitem[{Naidoo {et~al.}(2020)Naidoo, Whiteway, Massara, Gualdi, Lahav, \&
  Viel}]{Naidoo2019}
Naidoo, K., Whiteway, L., Massara, E., {et~al.} 2020, Monthly Notices of the
  Royal Astronomical Society, 491, 1709

\bibitem[{Paillas {et~al.}(2021)Paillas, Cai, Padilla, \&
  S{\'{a}}nchez}]{Paillas2021}
Paillas, E., Cai, Y.~C., Padilla, N., \& S{\'{a}}nchez, A.~G. 2021, Monthly
  Notices of the Royal Astronomical Society, 505, 5731

\bibitem[{Paillas {et~al.}(2022)Paillas, Cuesta-Lazaro, Zarrouk, Cai, Percival,
  Nadathur, Pinon, de~Mattia, \& Beutler}]{Paillas2022}
Paillas, E., Cuesta-Lazaro, C., Zarrouk, P., {et~al.} 2022, arXiv e-prints, 19,
  1

\bibitem[{Peacock {et~al.}(2001)Peacock, Cole, Norberg, Baugh, Bland-Hawthorn,
  Bridges, Cannon, Colless, Collins, Couch, Dalton, Deeley, {De Propris},
  Driver, Efstathiou, Ellis, Frenk, Glazebrook, Jackson, Lahav, Lewis, Lumsden,
  Maddox, Percival, Peterson, Price, Sutherland, \& Taylor}]{Peacock2001}
Peacock, J.~A., Cole, S., Norberg, P., {et~al.} 2001, Nature, 410, 169

\bibitem[{Philcox \& Ivanov(2022)}]{Philcox2022}
Philcox, O.~H. \& Ivanov, M.~M. 2022, Physical Review D, 105, 43517

\bibitem[{Sousbie(2011)}]{DisperseTheory}
Sousbie, T. 2011, Monthly Notices of the Royal Astronomical Society, 414, 350

\bibitem[{Spergel {et~al.}(2015)Spergel, Gehrels, Baltay, Bennett,
  Breckinridge, Donahue, Dressler, Gaudi, Greene, Guyon, Hirata, Kalirai,
  Kasdin, Macintosh, Moos, Perlmutter, Postman, Rauscher, Rhodes, Wang,
  Weinberg, Benford, Hudson, Jeong, Mellier, Traub, Yamada, Capak, Colbert,
  Masters, Penny, Savransky, Stern, Zimmerman, Barry, Bartusek, Carpenter,
  Cheng, Content, Dekens, Demers, Grady, Jackson, Kuan, Kruk, Melton, Nemati,
  Parvin, Poberezhskiy, Peddie, Ruffa, Wallace, Whipple, Wollack, \&
  Zhao}]{WFIRST15}
Spergel, D., Gehrels, N., Baltay, C., {et~al.} 2015, {Wide-Field InfrarRed
  Survey Telescope-Astrophysics Focused Telescope Assets WFIRST-AFTA 2015
  Report}, Tech. rep.

\bibitem[{Stoica {et~al.}(2007)Stoica, Mart{\'{i}}nez, \& Saar}]{Stoica2007}
Stoica, R.~S., Mart{\'{i}}nez, V.~J., \& Saar, E. 2007, Journal of the Royal
  Statistical Society. Series C: Applied Statistics, 56, 459

\bibitem[{Takahashi {et~al.}(2010)Takahashi, Yoshida, Takada, Matsubara,
  Sugiyama, Kayo, Nishimichi, Saito, \& Taruya}]{Takahashi2010}
Takahashi, R., Yoshida, N., Takada, M., {et~al.} 2010, Astrophysical Journal,
  726, 7

\bibitem[{Tempel {et~al.}(2014)Tempel, Stoica, Mart{\'{i}}nez, Liivam{\"{a}}gi,
  Castellan, \& Saar}]{Tempel2014}
Tempel, E., Stoica, R.~S., Mart{\'{i}}nez, V.~J., {et~al.} 2014, Monthly
  Notices of the Royal Astronomical Society, 438, 3465

\bibitem[{Totsuji \& Kihara(1969)}]{Totsuji1969}
Totsuji, H. \& Kihara, T. 1969, The Astronomical Society of Japan, 21, 221

\bibitem[{Uhlemann {et~al.}(2020)Uhlemann, Friedrich, Villaescusa-Navarro,
  Banerjee, \& Codis}]{Uhlemann2020}
Uhlemann, C., Friedrich, O., Villaescusa-Navarro, F., Banerjee, A., \& Codis,
  S. 2020, Monthly Notices of the Royal Astronomical Society, 495, 4006

\bibitem[{Valogiannis \& Dvorkin(2022{\natexlab{a}})}]{Valogiannis2022}
Valogiannis, G. \& Dvorkin, C. 2022{\natexlab{a}}, arXiv e-prints
  [\eprint[arXiv]{2204.13717}]

\bibitem[{Valogiannis \& Dvorkin(2022{\natexlab{b}})}]{Valogiannis2021}
Valogiannis, G. \& Dvorkin, C. 2022{\natexlab{b}}, Physical Review D, 105
  [\eprint[arXiv]{2108.07821}]

\bibitem[{Villaescusa-Navarro {et~al.}(2018)Villaescusa-Navarro, Banerjee,
  Dalal, Castorina, Scoccimarro, Angulo, \& Spergel}]{Villaescusa-Navarro2018}
Villaescusa-Navarro, F., Banerjee, A., Dalal, N., {et~al.} 2018, The
  Astrophysical Journal, 861, 53

\bibitem[{Villaescusa-Navarro {et~al.}(2020)Villaescusa-Navarro, Hahn, Massara,
  Banerjee, Delgado, Ramanah, Charnock, Giusarma, Li, Allys, Brochard,
  Uhlemann, Chiang, He, Pisani, Obuljen, Feng, Castorina, Contardo, Kreisch,
  Nicola, Alsing, Scoccimarro, Verde, Viel, Ho, Mallat, Wandelt, \&
  Spergel}]{Villaescusa-Navarro2019}
Villaescusa-Navarro, F., Hahn, C., Massara, E., {et~al.} 2020, The
  Astrophysical Journal Supplement Series, 250, 2

\bibitem[{Vogelsberger {et~al.}(2014)Vogelsberger, Genel, Springel, Torrey,
  Sijacki, Xu, Snyder, Nelson, \& Hernquist}]{Vogelsberger2014}
Vogelsberger, M., Genel, S., Springel, V., {et~al.} 2014, Monthly Notices of
  the Royal Astronomical Society, 444, 1518

\bibitem[{Wang \& He(2022)}]{Wang2022}
Wang, Y. \& He, P. 2022, The Astrophysical Journal, 934, 112

\bibitem[{Wang {et~al.}(2022)Wang, Yang, \& He}]{Wang2022a}
Wang, Y., Yang, H.-Y., \& He, P. 2022, The Astrophysical Journal, 934, 77

\bibitem[{Woodfinden {et~al.}(2022)Woodfinden, Nadathur, Percival, Radinovic,
  Massara, \& Winther}]{Woodfinden2022}
Woodfinden, A., Nadathur, S., Percival, W.~J., {et~al.} 2022, Monthly Notices
  of the Royal Astronomical Society, 516, 4307

\end{thebibliography}

\appendix

\section{Fisher formalism}   \label{appendix:Fisher}

The Fisher formalism allows the forecast of constraints we would get on our set of parameters $\theta = \{ \Omega_\mathrm{m}, \Omega_\mathrm{b}, h, n_\mathrm{s}, \sigma_8, M_\nu\}$ based on a statistical summary of the density fields. In our case, this summary is given by the set of (cross- or auto-) environment-dependent power spectra in Fourier bins, $\bm{s} = \{ P_{\alpha\beta}(k) \}$. Assuming a Gaussian likelihood with a covariance matrix independent from cosmology \citep{Carron2013, Kodwani2019}, we can derive elements of the Fisher information matrix as
\begin{equation} \label{eq:FIM_gaussian}
    \left[ \bm{I}(\bm{\theta}) \right]_{i,j} = \left(\frac{\partial \bar{\bm{s}}}{\partial \theta_i}\right)\tran \bm{\Sigma}^{-1} \left(\frac{\partial \bar{\bm{s}}}{\partial \theta_j}\right).
\end{equation}

The first ingredient of the computation of the Fisher information matrix is the unbiased estimate of the inverse covariance matrix $\bm{\Sigma}^{-1}$, which is given, still under the Gaussian assumption, by \citep{Kaufman1967, Hartlap2007}
\begin{equation} \label{eq:precision}
    \bm{\Sigma}^{-1} = \frac{N_\mathrm{fid} - n - 2}{N_\mathrm{fid} - 1} \, \hat{\bm{\Sigma}}^{-1},
\end{equation}
with $N_\mathrm{fid}$ the number simulations at the fiducial cosmology, $n$ is the length of the summary statistics vector $\bm{s}$ and $\hat{\bm{\Sigma}}=\left(\bm{s} - \bar{\bm{s}} \right) \left(\bm{s} - \bar{\bm{s}} \right)\tran / \left(N_\mathrm{fid}-1\right)$ is the unbiased estimate of the covariance matrix.

The second ingredient of Eq.~\eqref{eq:FIM_gaussian} are the derivatives with respect to the parameters of the model $\theta$. These latter can be estimated numerically from the $N_\mathrm{deriv} = 500$ realisations of variation of individual parameter in the Quijote suite as
\begin{equation} \label{eq:derivatives}
    \frac{\partial \bar{\bm{s}}}{\partial \theta_i} \simeq \frac{\bar{\bm{s}}(\theta_i + \dd \theta_i) - \bar{\bm{s}}(\theta_i - \dd \theta_i)}{2 \dd \theta_i}. 
\end{equation}
We note however that such a definition do not apply for $M_\nu$ being a positive quantity with a fiducial value at $0.0$ eV. For this parameter, we thus rely on the four-point forward approximation of the derivative
\begin{equation} \label{eq:derivatives_neutrinos}
    \frac{\partial \bar{\bm{s}}}{\partial M_\nu} \simeq \frac{ \bar{\bm{s}}(4 M_\nu^+) - 12 \bar{\bm{s}}(2 M_\nu^+) + 32 \bar{\bm{s}}(M_\nu^+) - 21\bar{\bm{s}}(M_\nu = 0.0)}{12 M_\nu^+},
\end{equation}
where $M_\nu^+ = 0.1$ eV. For consistency, we use a set of simulations initialised using the Zel'dovich approximation to compute $\bar{\bm{s}}(M_\nu = 0)$ in the previous equation since all the other terms are initialised this way. The presented results are obtained making use of $N_\mathrm{deriv}=500$ and $N_\mathrm{fid}=7000$ realisations to respectively compute the derivatives and the covariance matrix.

\section{Covariance matrices, derivatives, and confidence ellipses} \label{appendix:cov_deriv}

\begin{figure}
        \centering
        \includegraphics[width=1\linewidth]{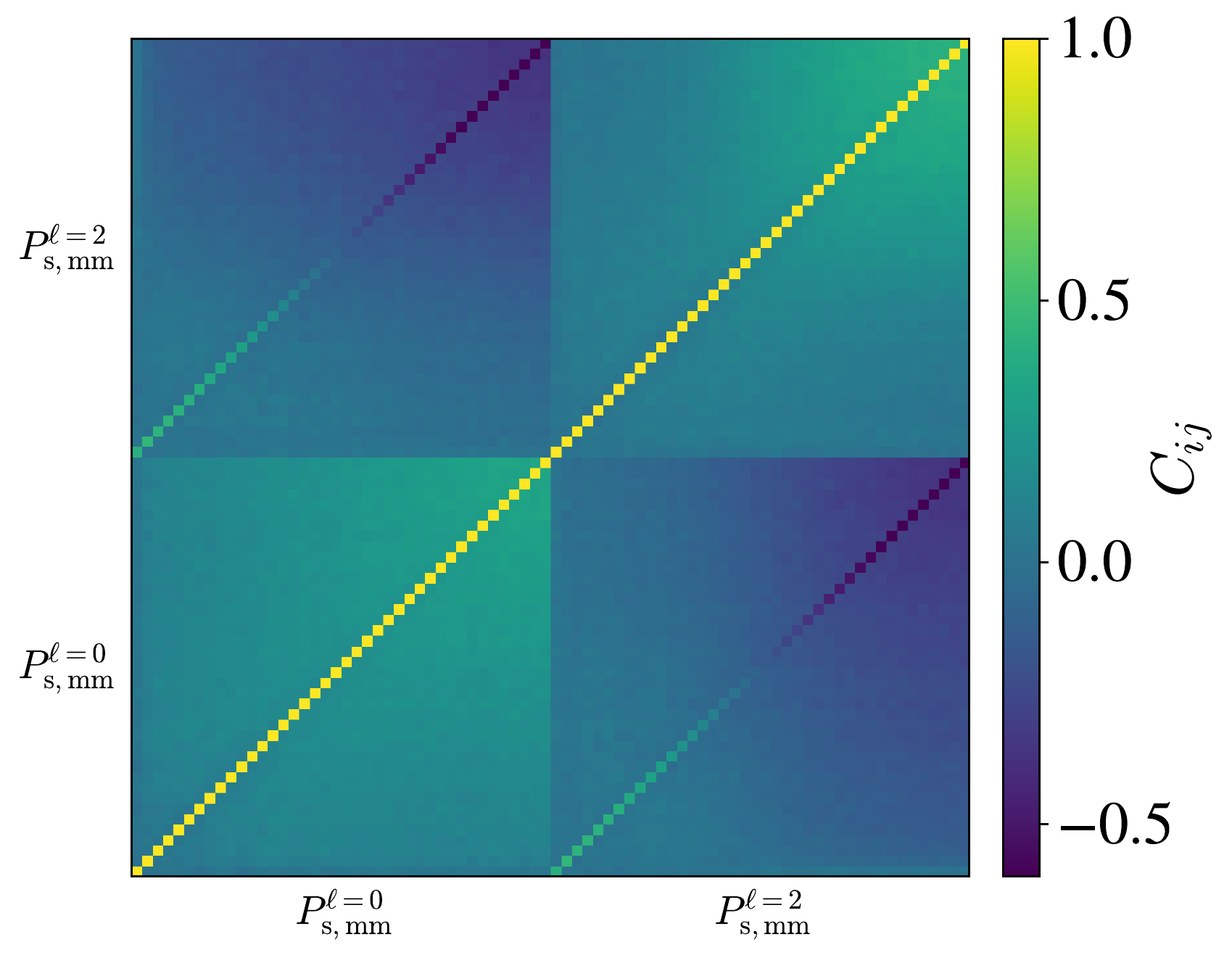}
        \includegraphics[width=1\linewidth]{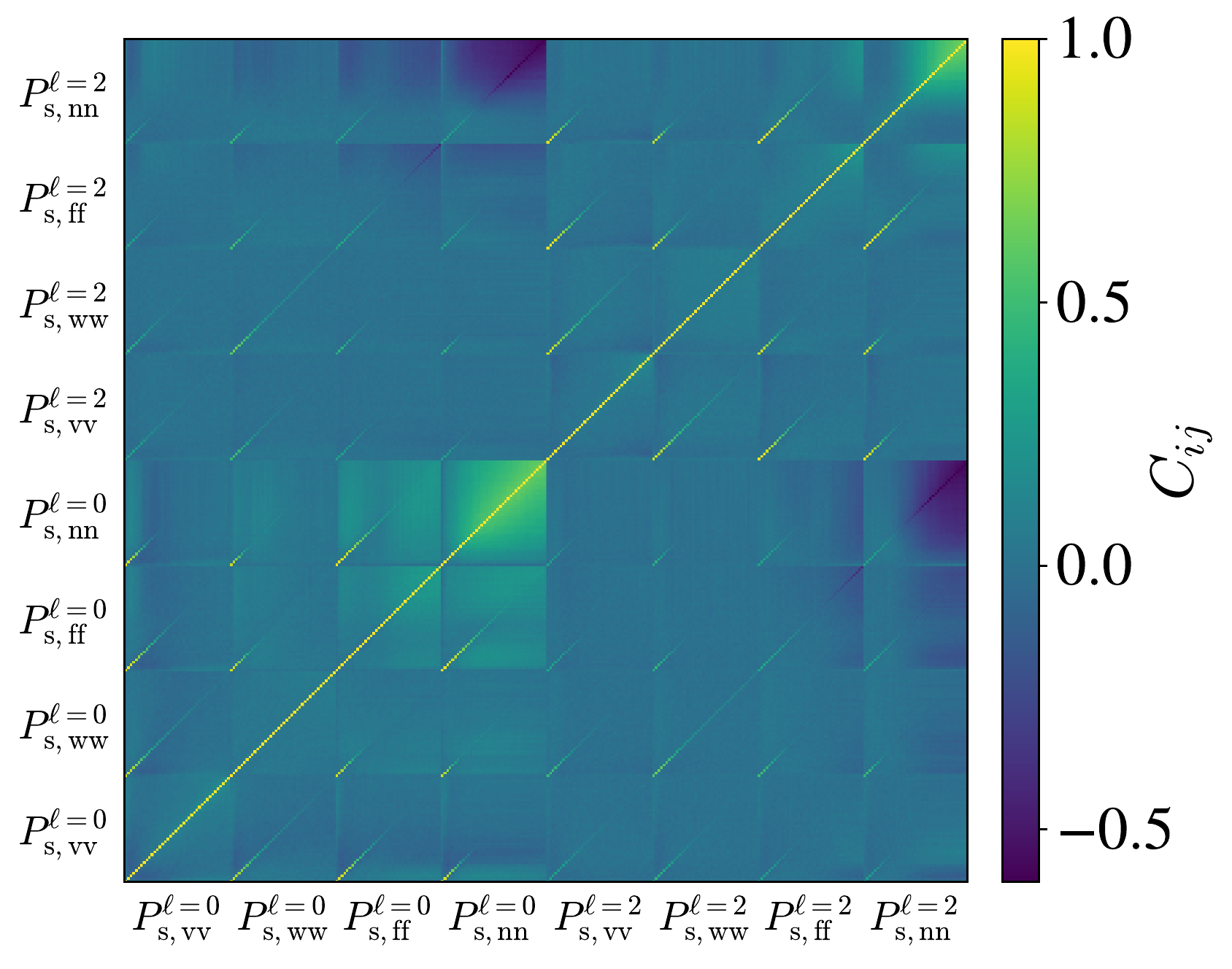}
        \caption{\textit{(top)} Correlation coefficients $C_{ij}$ for the matter power spectrum. \textit{(bottom)} Same for $P_{\alpha\alpha}(k)$ coefficients extracted from the several environments. Each sub-matrix goes from $k=0.1$ $h$/Mpc to $k=0.5$ $h$/Mpc.}
        \label{fig:correlation}
    \end{figure}

\begin{figure*}
        \centering
        \includegraphics[width=1\linewidth]{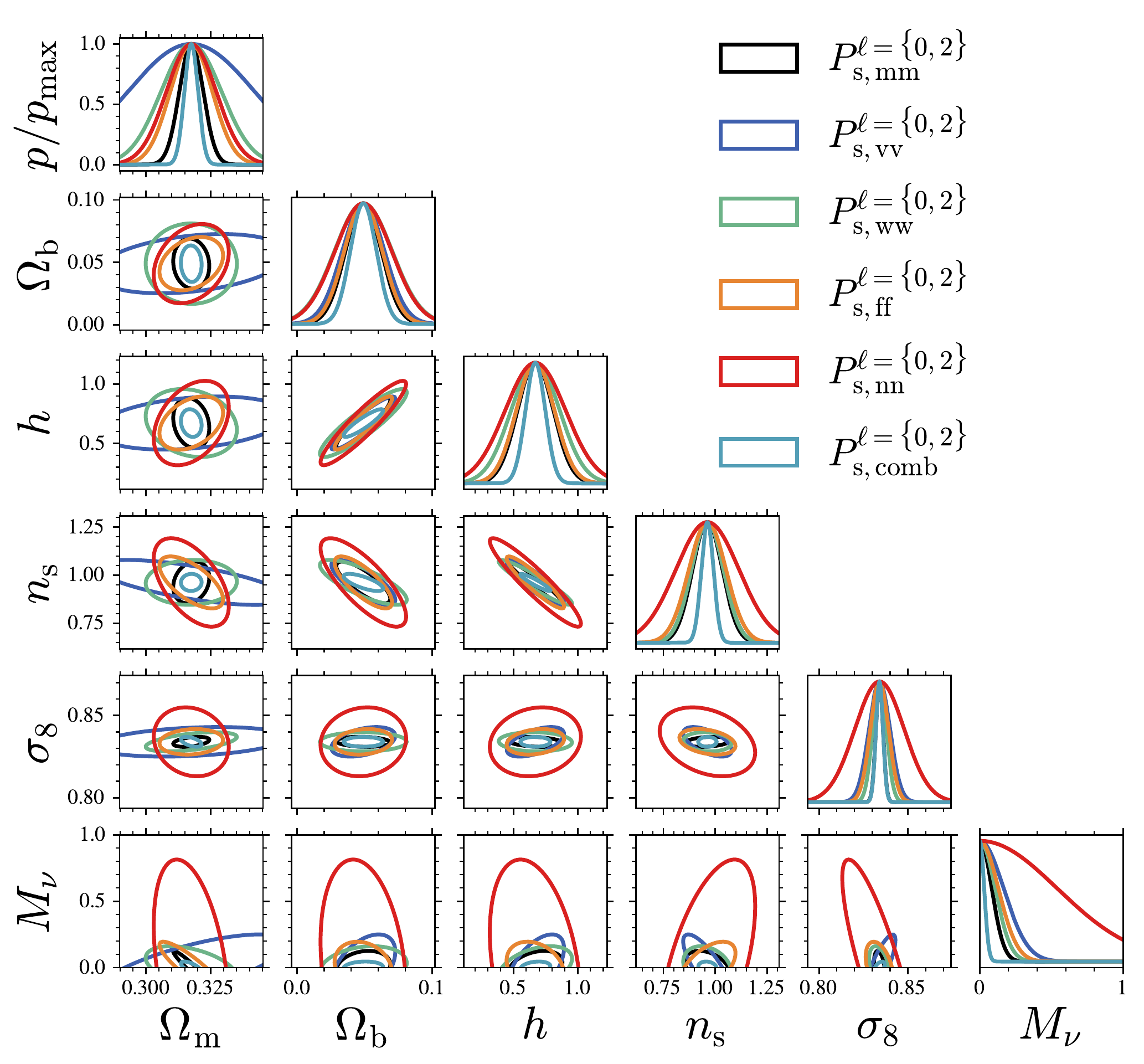}
        \caption{$1\sigma$ confidence ellipses for all the pairs of cosmological parameters $\left( \Omega_\mathrm{m}, \Omega_\mathrm{b}, h, n_\mathrm{s}, \sigma_\mathrm{8}, M_\nu \right)$ obtained from the several statistics (either the matter monopole+quadrupole or the ones from the environments and their combination) in redshift space. On the diagonal are shown the normalised probability density functions for each parameter.}
        \label{fig:corner}
    \end{figure*}

    While discussed in the main text Sect. \ref{sect:constraints}, we report in this section the two ingredients of the Fisher quantification of information that are: (i) the correlation matrices between the several elements of the statistic vector in Fig.~\ref{fig:correlation} and (ii) the derivatives of the redshift-space monopoles and quadrupoles with respect to the six cosmological parameters.
    Figure \ref{fig:corner} shows the corner plot with the several confidence ellipses obtained for the $P^{\ell=\{0, 2\}}_\mathrm{s, \alpha\alpha}$ statistics for all the six cosmological parameters $\Omega_\mathrm{m}$, $\Omega_\mathrm{b}$, $h$, $n_\mathrm{s}$, $\sigma_8$, $M_\nu$.
    
\section{Constraints from spectra in the several environments}

We provide here the individual constraints obtained from the monopole and quadrupole power spectra computed in the different cosmic web environments in Table~\ref{tab:constraints_envs}. As discussed in Sect.~\ref{sect:constraints}, they all are leading to much weaker constraints than the full matter statistic in redshift-space when $k_{\mathrm{max}} = 0.5$ $h$/Mpc. However, their combination leads to tighter constraints with an improvement by a factor of up to $2.7$ for the summed neutrino mass and $2.4$ for $n_\mathrm{s}$.

\begin{table*}
    \centering
    \caption{Marginalised 1-$\sigma$ constraints obtained from the analysis of power spectra monopoles and quadrupoles computed in the different environments for all cosmological parameters. All the improvement factors are relative to the matter case in redshift-space, namely $P^{\ell = \{0, 2\}}_\mathrm{s, mm}$ given in the third row of Table~\ref{tab:constraints_RSD_with_quad}. $\sigma_{M_\nu}$ is in unit of eV.}
    \label{tab:constraints_envs}
    
    \renewcommand{\arraystretch}{1.5}
    \smallskip
    \begin{adjustbox}{max width=1.0\textwidth,center}
    \begin{tabular}{c|cccccc}
        \toprule
        Statistics & $\sigma_{\Omega_\mathrm{m}}$ & $\sigma_{\Omega_\mathrm{b}}$ & $\sigma_{h}$ & $\sigma_{n_\mathrm{s}}$ & $\sigma_{\sigma_\mathrm{8}}$ & $\sigma_{M_\nu}$ \\
        
        \midrule
        
        $\color{void} P_\mathrm{s,vv}^{\ell=\{0,2\}}$ & $0.0244 \, (0.2)$ & $0.0157 \, (0.8)$ & $0.1478 \, (0.9)$ & $0.0773 \, (0.9)$ & $0.0060 \, (0.3)$ & $0.1657 \, (0.5)$ \\ 
        
        $\color{wall} P_\mathrm{s,ww}^{\ell=\{0,2\}}$ & $0.0117 \, (0.4)$ & $0.0213 \, (0.6)$ & $0.1893 \, (0.7)$ & $0.0772 \, (0.9)$ & $0.0039 \, (0.5)$ & $0.1069 \, (0.8)$ \\
        
        $\color{filament} P_\mathrm{s,ff}^{\ell=\{0,2\}}$ & $0.0081 \, (0.6)$ & $0.0141 \, (0.9)$ & $0.1485 \, (0.9)$ & $0.0895 \, (0.8)$ & $0.0051 \, (0.4)$ & $0.1294 \, (0.6)$ \\
        
        $\color{node} P_\mathrm{s,nn}^{\ell=\{0,2\}}$ & $0.0096 \, (0.5) $ & $0.0210 \, (0.6) $ & $0.2350 \, (0.6)$ & $0.1523 \, (0.5)$ & $0.0139 \, (0.1)$ & $0.5397 \, (0.2)$ \\
        
        $\color{combination} P_\mathrm{s, comb}^{\ell=\{0,2\}}$ & $0.0027 \, (1.7)$ & $0.0097 \, (1.4)$ & $0.0773 \, (1.8)$ & $0.0295 \, (2.4)$ & $0.0020 \, (1)$ & $0.0304 \, (2.7)$ \\
        
        \bottomrule
    \end{tabular}
    \end{adjustbox}
    \end{table*}

\end{document}